\documentclass{bmcart}

\usepackage[utf8]{inputenc} 

\usepackage[]{algorithm2e}
\usepackage{array}
\usepackage{amssymb}
\usepackage{amsmath}
\usepackage{graphicx}
\usepackage{hyperref}
\usepackage{wasysym}

\startlocaldefs
\endlocaldefs

\begin{document}

\begin{frontmatter}

\begin{fmbox}
\dochead{Research}


\title{Crisis contagion in the world trade network}


\author[
   addressref={aff1},                   
   email={celestin.coquide@utinam.cnrs.fr}   
]{\inits{C}\fnm{C\'elestin} \snm{Coquid\'e}}
\author[
   addressref={aff1},
   corref={aff1},                       
   email={jose.lages@utinam.cnrs.fr}
]{\inits{J}\fnm{Jos\'e} \snm{Lages}}
\author[
addressref={aff2},
email={dima@irsamc.ups-tlse.fr}
]{\inits{DL}\fnm{Dima L} \snm{Shepelyansky}}


\address[id=aff1]{
  \orgname{Institut UTINAM, UMR 6213, CNRS, Universit\'e Bourgogne Franche-Comt\'e}, 
  \city{Besan\c con},                              
  \cny{France}                                    
}
\address[id=aff2]{%
  \orgname{Laboratoire de Physique Th\'eorique, IRSAMC, Universit\'e de Toulouse, CNRS, UPS},
  \city{Toulouse},
  \cny{France}
}



\end{fmbox}


\begin{abstractbox}

\begin{abstract} 
%
We present a model of worldwide crisis contagion based on the Google matrix analysis of the world trade network obtained from the UN Comtrade database. The fraction of bankrupted countries exhibits an \textit{on-off} phase transition governed by a bankruptcy threshold $\kappa$ related to the trade balance of the countries. For $\kappa>\kappa_c$, the contagion is circumscribed to less than 10\% of the countries, whereas, for $\kappa<\kappa_c$, the crisis is global with about 90\% of the countries going to bankruptcy. We measure the total cost of the crisis during the contagion process. In addition to providing contagion scenarios, our model allows to probe the structural trading dependencies between countries. For different networks extracted from the world trade exchanges of the last two decades, the global crisis comes from the Western world. In particular, the source of the global crisis is systematically the Old Continent and The Americas (mainly US and Mexico). Besides the economy of Australia, those of Asian countries, such as China, India, Indonesia, Malaysia and Thailand, are the last to fall during the contagion. Also, the four BRIC are among the most robust countries to the world trade crisis.
\end{abstract}




\end{abstractbox}
%

\end{frontmatter}



\section*{Introduction}

The financial crisis of 2007-2008 highlighted the enormous effect of contagion 
over world bank networks (see e.g. \cite{gai10,elliott14,fink16}). Similar contagion effects appear also in the world trade
which is especially vulnerable to energy crisis mainly related to the trade of petroleum and gas (see e.g. \cite{energy,oil}).
In this work, we model the crisis contagion in the world trade 
using the UN Comtrade database \cite{comtrade}. We use the Google matrix analysis \cite{brin98,langville12,ermann15b} of the world trade network (WTN)
developed in \cite{ermann11,ermann15}. In comparison with the usual import-export analysis based on the counting of trade volumes directly exchanged between countries, the advantage of the Google matrix analysis is that the long range interactions between the network nodes, i.e., the countries, are taken into account. Otherwise stated, this analysis captures the fact that even two countries which are not direct trade partners can possibly have their economies correlated through the cascade of trade exchanges between a chain of intermediary countries. 
The power of the specific Google matrix related algorithms, such as the PageRank algorithm, is well illustrated by the success of the Google search engine \cite{brin98,langville12}, and also by their possible applications to a rich variety of directed networks (see \cite{ermann15b} for a review). The detailed UN Comtrade database, collected for about 50 years,
allows to perform a thorough modeling of the
crisis contagion in the WTN. In the following, we use the contagion model
inspired by the analysis of the crisis in the Bitcoin transactions network presented in \cite{ermann18,coquide19c}.

We note that various research  groups
studied the statistical properties
of the world trade network
(see e.g. \cite{serrano07,fagiolo09,he10,fagiolo10,barigozzi10,debenedictis11,deguchi14})
but the contagion process has not been analyzed so far.
We think that our study will attract research interest
to this nontrivial and complex process.
Such an analysis can be also extended to
networks of interconnected banks (see e.g. \cite{roncorini14})
where the contagion process is of primary importance.

\section*{Datasets}
\begin{figure}[t!]
	\centering
	\includegraphics[width=0.95\columnwidth]{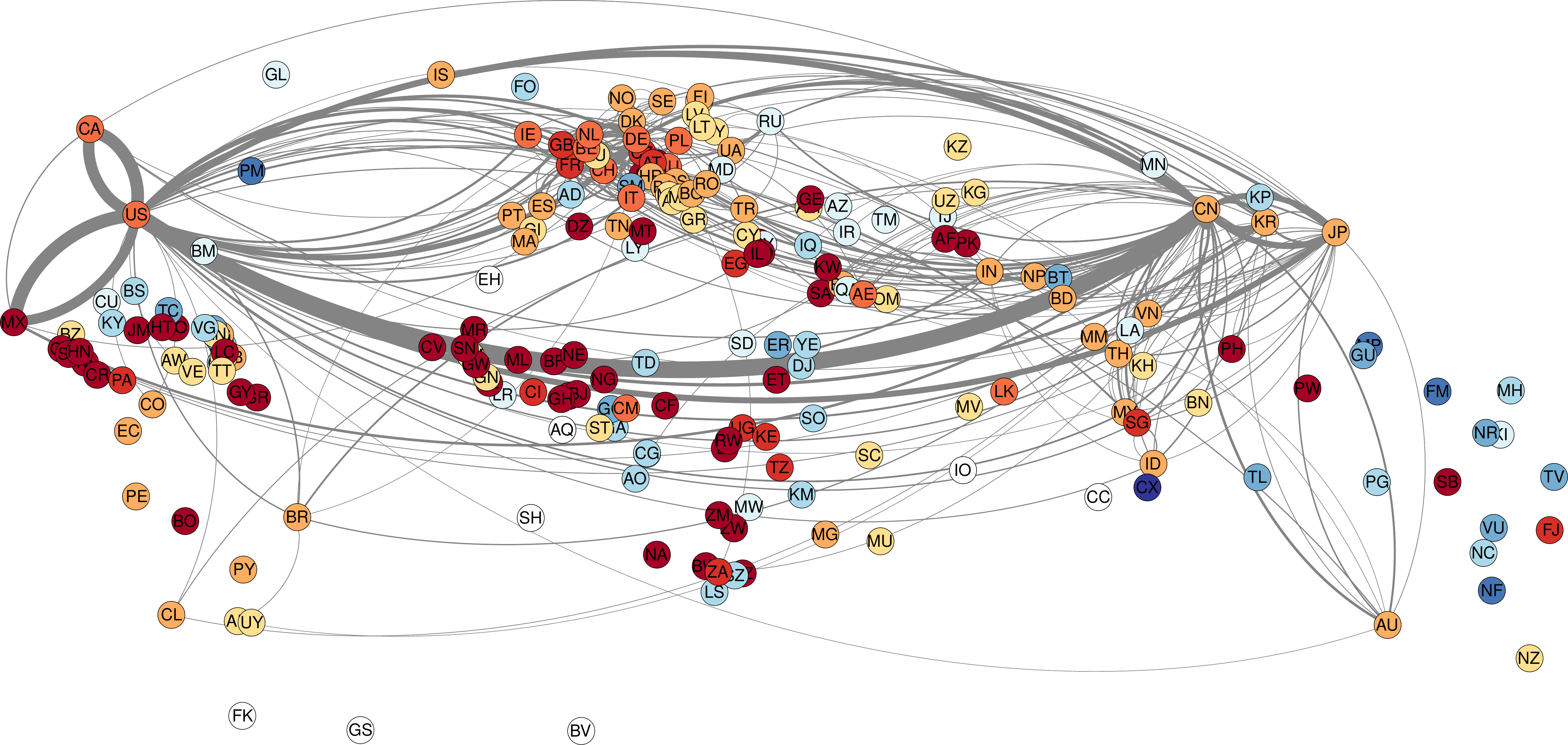}
	\caption{\csentence{2016 World trade network.}
		Two countries A and B are related by a directed link, the direction of which is given by its curvature. If A points to B following the bent path in the clockwise direction (A$\frown$B) then A exports to B, otherwise, i.e. (A$\smile$B), B exports to A. The width of the link is proportional to the exportation volume in the WTN from the source country to the target country.
		The colors of country nodes range from red (blue) for a country going to bankruptcy at stage $\tau=0$ ($\tau=\tau_{\infty}$) in the case of a bankruptcy threshold $\kappa=0.1$ and for the following crisis scenario: once a country goes to bankruptcy, it is prevented to import products with the exception of petroleum and gas (see details in the \nameref{sec:contagion} section).
		Only transactions above $10^{10}$~USD are shown. Most of the Polynesian islands have been removed, here and in the following figures, to improve visibility.}
	\label{fig1}
\end{figure}

Using the UN Comtrade database \cite{comtrade}, we construct the multiproduct World Trade Network (WTN) for the years 2004, 2008, 2012 and 2016. Each year is characterized by a money matrix, $M^{p}_{cc^{\prime}}$, giving the export flow of product $p$, expressed in USD, from country $c^{\prime}$ to country $c$. The data concern a set $\mathcal{C}$ of $N_{c} = 227$ countries and territories, and a set $\mathcal{P}$ of $N_{p} = 61$ principal type  of products. The list of these products, which belongs to the Standard International Trade Classification (Rev. 1), is given in \cite{ermann15}. The 2016 WTN is represented in Fig.~\ref{fig1}.
The set $\mathcal{C}$ comprises $N_c=227$ sovereign states and territories which are listed, with their associated ISO 3166-1 alpha-2 code, in the \nameref{sec:abbreviation} section. Among territories, most of them belong to a sovereign state, some are disputed territories, such as Western Sahara, and Antarctica is a international condominium. The UN Comtrade database inventories commodities flows, not only between sovereign states, but also from and to these territories.
The present study complies with the UN Comtrade terms of use.

\section*{Model}
In this section, we recall the construction process of the google matrix $G$ associated to the WTN, and the PageRank-CheiRank trade balance (PCTB) \cite{ermann15,coquide19,coquide19b}. We introduce also a model of crisis contagion in the WTN.

\subsection*{Multiproduct World Trade Network}

For a given year, the multiproduct WTN is characterized by $N_cN_p$ nodes, each one representing a couple of country and product ($cp$). We assign a weight $M_{cc'}^p$ to the directed link from node $c'p$ to $cp$. We define $V_{cp}=\sum_{c'}M_{cc'}^p$
as the total volume of product $p$ imported by the country $c$, and $V^*_{cp}=\sum_{c'}M_{c'c}^p$ as the total volume of product $p$ exported from the country $c$.

\subsection*{Google matrix of the World Trade Network}
The Google matrix $G$ is constructed as
\begin{equation}
G=\alpha S + (1- \alpha)\mathbf{v} \mathbf{e^{T}}
\label{eq:G}
\end{equation}
where $S$ is a stochastic matrix, the elements of which are
\begin{equation}
S_{cp,c'p'}=\left\{
\begin{array}{cl}
\delta_{pp'}M^p_{cc'}/V^{*}_{c'p}&\text{if }V^{*}_{c'p}\neq 0\\
1/N&\text{if }V^{*}_{c'p}=0
\end{array}\right..
\label{eq:S}
\end{equation}
Here, $\alpha\in\left[0.5,1\right[$ is the damping factor,
$\mathbf{v}$ is a preferential probability vector,
and
$\mathbf{e^{T}}=(1,1,\dots,1)$ is a row vector.
The Google matrix $G$ (\ref{eq:G}) describes the transition probabilities of a random surfer which, with a probability $\alpha$, follows the architecture of the multiproduct WTN encoded in the stochastic matrix $S$, and, with a probability $(1-\alpha)$, jumps to any node of the WTN according to the preferential probability vector $\bf v$. Below, we use either $\alpha=0.5$ or $\alpha=0.85$. This second value is the one used in the seminal paper of Brin and Page devoted to the PageRank algorithm \cite{brin98}.
The PageRank vector $\bf P$ characterizes the steady state of the Markovian process described by the Google matrix $G$ (\ref{eq:G}), i.e., $G{\bf P}={\bf P}$. The $cp$ component of the PageRank vector $\bf P$, i.e., $P_{cp}$, gives the fraction of time the random surfer spent on the node $cp$ during its infinite journey in the WTN.

Following \cite{ermann15,coquide19}, the final WTN Google matrix is obtained after two contruction steps. We use a first preferential probability vector ${\bf v}^1$, the components of which are $v^1_{cp}=V_{cp}/(N_{c}V_c)$ where $V_c=\sum_{p}V_{cp}$ is the total volume of commodities imported by the country $c$. This choice of the preferential probability vector ensures equity for the random jumps between countries. This preferential probability vector ${\bf v}^1$ allows to compute the PageRank vector ${\bf P}^1$ associated to the Google matrix $G^1$. As a second step, we use the PageRank vector ${\bf P}^1$ to define a new preferential probability vector $\bf v$, the components of which are 
$v_{cp}=P^1_p/N_c$ where $P^1_p=\sum_{c'} P^{1}_{c'p}$ gives the ability of a product $p$ to be imported.
The final Google matrix $G$ (\ref{eq:G}) is constructed using the latter defined preferential probability vector $\mathbf{v}$. The PageRank vector component $P_{cp}$ naturally characterizes the ability of a country $c$ to import a product $p$ \cite{ermann15,coquide19}.

It is interesting to consider the complex network built by inverting the directed links of the WTN. The Google matrix $G^*$ associated to this inverted network is obtained from the stochastic matrix $S^*$, the elements of which are
\begin{equation}
S_{cp,c'p'}=\left\{
\begin{array}{cl}
\delta_{pp'}M^p_{c'c}/V_{c'p}&\text{if }V_{c'p}\neq 0\\
1/N&\text{if }V_{c'p}=0
\end{array}\right.,
\label{eq:Sstar}
\end{equation}
and from the preferential probability vectors ${\bf v}^{*1}$ and ${\bf v}^*$, the components of which are 
$v^{*1}_{cp}=V^*_{cp}/(N_{c}V^*_c)$
and
$v^*_{cp}=P^{*1}_p/N_c$, where $V^*_c=\sum_pV^*_{cp}$ is the total export volume of the country $c$ and $P^{*1}_p=\sum_{c'} P^{*1}_{c'p}$ gives the ability of the product $p$ to be exported. Here, ${\bf P}^{*1}$ and ${\bf P}^*$ are the CheiRank vectors defined such as $G^{*1}{\bf P}^{*1}={\bf P}^{*1}$ and $G^*{\bf P}^*={\bf P}^*$.
The CheiRank vector component $P^*_{cp}$ naturally characterizes the ability of a country $c$ to export a product $p$ \cite{ermann15,coquide19,coquide19b}.

In addition to the PageRank vector $\bf P$ and the CheiRank vector ${\bf P}^*$, we can define the ImportRank vector $\bf I$ and the ExportRank vector $\bf E$, the components of which are
$I_{cp}=V_{cp}/V$ and $E_{cp}=V^*_{cp}/V$
where $V$ is the total volume exchanged through the WTN.
The ImportRank and ExportRank constitute crude accounting measures of the capabilities of a country $c$ to import or export a given product $p$. It has been shown \cite{ermann15,coquide19b} that the rankings by PageRank and CheiRank provide a more finer measure of these capabilities since it takes account of the all the direct ($c'p\rightarrow cp$) and indirect ($c'p\rightarrow c_1p\rightarrow c_2p\rightarrow \dots\rightarrow cp$) economical exchanges of any commodity $p$ between any pair of countries $c'$ and $c$. The PageRank and CheiRank algorithms express the economical importance of a $(cp)$-pair, i.e., a country-product pair, inside the complex network constituted by the international trade.

\subsection*{PageRank-CheiRank trade balance}
As the PageRank and CheiRank algorithms measure the capabilities of a country to import or to export products, we can define the PageRank-CheiRank trade balance (PCTB) of a given country $c$ as
\begin{equation}\label{eq:PCTB}
B_c=\displaystyle\frac{P^*_c-P_c}{P^*_c+P_c}
\end{equation}
where $P_c=\sum_pP_{cp}$ is the country $c$ PageRank component and $P^*_c=\sum_pP^*_{cp}$ the country $c$ CheiRank component. The PCTB is bounded, $B_c\in\left[-1,1\right]$. The more $B_c$ is positive, the more the country $c$ is a more efficient exporter than importer in the WTN. Consequently, the country $c$ economic health should be correlated with the value of $B_c$. 

Analogously, the usual normalized import-export trade balance can be defined using the Import\-Rank and the ExportRank as
\begin{equation}\label{eq:IEB}
\hat{B}_c=\displaystyle\frac{E_c-I_c}{E_c+I_c}
\end{equation}
where
$E_c=\sum_{p}E_{cp}$ is the country $c$ total export volume (divided by $V$)
and where
$I_c=\sum_{p}I_{cp}$ is the country $c$ total import volume (divided by $V$).

\subsection*{Contagion model}
\label{sec:contagion}

Countries with large negative PCTB naturally have to restrain their imports of non vital goods. This restriction can be \textit{de facto}, as not enough liquidity are available for these countries, or can be imposed by a supranational organization  in order to hold back a possible crisis contagion (e.g. the European Union for countries belonging to the Eurozone).
Thus, let us assume that every country $c$ with $B_c\leq-\kappa$ goes to bankruptcy. Here, $\kappa\geq0$ is the bankruptcy threshold.

\begin{algorithm}[t]\label{algo1}
	\SetKwInOut{Data}{data}
	\SetKwInOut{Input}{input}\SetKwInOut{Output}{output}
	\Data{WTN money matrix $M$}
	\Input{Bankruptcy threshold $\kappa$}
	\Output{Countries went in bankruptcy $\mathcal{B}_\tau$ at the crisis stage $\tau$}
	$\tau=0$, $\mathcal{B}_{-1}=\emptyset$
	
	\Repeat{$\mathcal{B}_\tau=\emptyset$}{
		$\mathcal{B}_\tau=\emptyset$\\
		Using $M$, compute $G$, $G^*$, $\mathbf{P}$ and $\mathbf{P}^*$
		
		\For{$c\in\mathcal{C}-\bigcup_{i=0}^{\tau-1}\mathcal{B}_i$}{
			\If{$B_c\leq-\kappa$}{
				$\mathcal{B}_\tau=\mathcal{B}_\tau+c$\\
				\If{model A}{
					\ForEach{$c'p\in\mathcal{C}\times\tilde{\mathcal{P}}|M_{cc'}^p\neq0$}{$M_{cc'}^p=0$}
				}
				\If{model B}{
					\ForEach{$c'p\in\mathcal{C}\times\mathcal{P}|M_{cc'}^p\neq0$}{$M_{cc'}^p=0$}
				}
			}
		}
		$\tau=\tau+1$
	}
	$\tau_{\infty}=\tau$
	
	\caption{Crisis contagion in the WTN.}
\end{algorithm}

At the crisis stage $\tau=0$, using the Google matrix $G_0=G$ defined by (\ref{eq:G}), we compute the PCTB $B_c$ for each country $c$. We obtain a set of countries $\mathcal{B}_0=\left\{c\in\mathcal{C}\,|\,B_c\leq-\kappa\right\}$ which go to bankruptcy at the crisis stage $\tau=0$ and which remain in this state in the following crisis stages $\tau>0$. Let us assume that all the bankrupted countries are prevented to import products at the following stages, $\tau\geq1$. We will consider two cases: the import ban concerns all the products with the exception of petroleum and gas (model A) or the import ban concerns all the products (model B). At the stage $\tau=1$, the world trade network is modified setting to zero the money matrix elements corresponding to the banned trade exchanges, i.e.,
\begin{equation}\label{eq:Mmod}
M_{cc'}^p=0 ,\forall c'\in\mathcal{C},\forall c\in\mathcal{B}_0,
\left\{
\begin{array}{l}
\forall p\in\tilde{\mathcal{P}}\qquad\mbox{(model A)}\\
\forall p\in\mathcal{P} \qquad\mbox{(model B)}
\end{array}
\right.
\end{equation}
where $\tilde{\mathcal{P}}=\mathcal{P}-\left\{\mbox{petroleum},\mbox{gas}\right\}$ is the set of all the exchanged commodities in the WTN with the exception of petroleum and gas.
The Google matrix $G_1$ is constructed using the above modified money matrix $M$ (\ref{eq:Mmod}). We compute again the PCTB for each country, and establish the set of countries, $\mathcal{B}_1=\left\{c\in\mathcal{C}-\mathcal{B}_0\,|\,B_c\leq -\kappa\right\}$, which go to bankruptcy at the stage $\tau=1$ and will remain in this state at later stages $\tau>1$, according to model A or model B. The crisis contagion stops at the contagion step $\tau_\infty$ for which no more countries go to bankruptcy. The WTN crisis contagion model is described by the Algorithm~\ref{algo1}.
This contagion model has already been used to analyze the crisis contagion in the bitcoin transaction network \cite{coquide19c}.

Let us define the proportion $\eta\left(\tau,\kappa\right)$ of the world countries in bankruptcy at the crisis stage $\tau$ for the bankruptcy threshold $\kappa$. Here, $\eta=0$ if no countries are in bankruptcy, and $\eta=1$ if all the $N_c$ countries and territories are in bankruptcy. For a given bankruptcy threshold $\kappa$, let us also define the cost of the crisis up to the end of the contagion stage $\tau$
\begin{equation}\label{eq:Cinf}
	C\left(\kappa,\tau\right)=\sum_{c\in\bigcup\limits_{\tau'=0}^{\tau}\mathcal{B}_{\tau'}}\sum\limits_{\scriptsize\begin{array}{c}
		p\in\tilde{\mathcal{P}}\\
		\mbox{(model A)}\\
		\mbox{or}\\
		p\in\mathcal{P}\\
		\mbox{(model B)}
		\end{array}} V_{cp}.
\end{equation}
The value of $C_\infty\left(\kappa\right)=C\left(\kappa,\tau_{\infty}\right)$ gives the total cost of the crisis, i.e., it gives the total volume of all the non accomplished commercial exchanges due to the successive bankruptcy of countries during the crisis contagion.

\begin{figure}[t!]
	\centering
	\includegraphics[width=0.95\columnwidth]{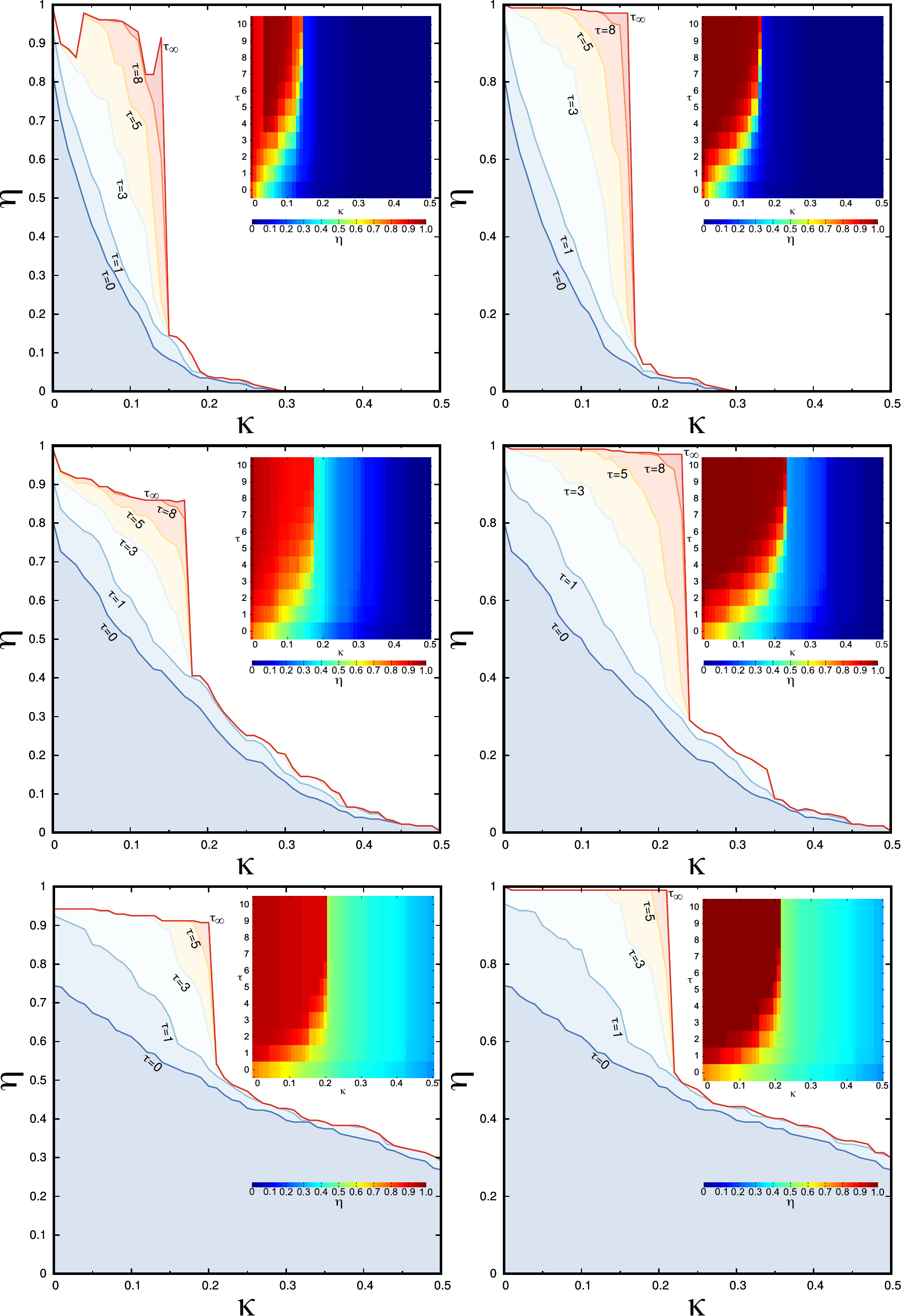}
	\caption{\csentence{Fraction of bankrupted countries for the 2016 WTN.} Fraction $\eta$ of countries went to bankruptcy up to the $\tau$th stage of the crisis contagion as a function of the bankruptcy threshold $\kappa$. For the first column, once a country goes to bankruptcy, it is prevented to import products with the exception of petroleum and gas (model A). For the second column, once a country goes to bankruptcy, it is prevented to import any product (model B). The first (second) row corresponds to a damping factor $\alpha=0.5$ ($\alpha=0.85$).
		The evolution of the fraction of bankrupted countries is monitored by the PCTB (\ref{eq:PCTB}) (first and second rows) and by the ImportRank-ExportRank balance (\ref{eq:IEB}) (third row).
		The insets show the corresponding fraction $\eta$ of bankrupted countries in the $\left(\tau,\kappa\right)$ plane. Dark red corresponds to the case where all the countries went to bankruptcy ($\eta=1$), and dark blue to the case where all the countries are safe ($\eta=0$).}
	\label{fig2}
\end{figure}

\section*{Results}

\subsection*{Phase transition of the crisis contagion}

The crisis contagion in the 2016 WTN is observed in Fig.~\ref{fig2} where the fraction $\eta$ of countries which go to bankruptcy is displayed as a function of the crisis contagion stage $\tau$ and of the bankruptcy threshold $\kappa$. We clearly see a transition from a regime of contained contagion for $\kappa>\kappa_c$ to a regime of global contagion for $\kappa<\kappa_{c}$. The Brin \& Page original damping factor value, i.e., $\alpha=0.85$, leads to a less frank transition (Fig.~\ref{fig2}, second row) than the $\alpha=0.5$ value which exhibits an ``all or nothing'' transition at $\kappa_{c}$ (Fig.~\ref{fig2}, first row). We note also that the critical bankruptcy threshold is, for $\alpha=0.5$, $\kappa_{c}\simeq0.15$ (model A) and $\kappa_c\simeq0.175$ (model B), and, for $\alpha=0.85$, $\kappa_{c}\simeq0.18$ (model A) and $\kappa_c\simeq0.24$ (model B). For a given bankruptcy threshold $\kappa$, the more $\alpha$ is low, the more the contagion is able to spread all over the WTN. This explain that for $\alpha=0.5$ the transition is more abrupt and the critical bankruptcy threshold $\kappa_c$ is lower than for $\alpha=0.85$. The model A is more realistic than the model B since a country in bankruptcy still needs to import vital commodities, as petroleum and gas, in order to support its industry which in return will provide commodities to export.
For low $\kappa$, the model B leads to a more global contagion crisis ($\eta\simeq1$) than the model A since the latter model indirectly protects countries which are petroleum and/or gas exporters.


Let us use the ImportRank and the ExportRank, and consequently, the normalized import-export trade balance $\hat{B}_c$ (\ref{eq:IEB}), to monitor the crisis contagion. In this case, the third row of Fig.~\ref{fig2} shows the fraction $\eta$ of countries in bankruptcy as a function of the bankruptcy threshold $\kappa$. We observe that for any $\kappa$, more than a third of the countries go to bankruptcy already at the $\tau=0$ crisis stage.
Moreover, there is no crisis containment for $\kappa>\kappa_{c}$ , since for a bankruptcy threshold $\kappa$ just above the critical value $\kappa_{c}$ half of the world countries and territories are in bankruptcy already at the stage $\tau=0$ of the contagion.
This ImportRank-ExportRank description is less suitable than the PageRank-CheiRank description to follow the crisis contagion since the transition around $\kappa_c\simeq0.2$ in Fig.~\ref{fig2} (third row) is less frank. Indeed, for model A, at the transition, $\eta$ goes from $0.5$ to $0.9$ (Fig.~\ref{fig2} third row, left column), while $\eta$ goes from $0.15$ to $0.9$ for the PageRank-CheiRank description (Fig.~\ref{fig2} first row, left column).
For a given country $c$, the ImportRank-ExportRank description is based only on the relative balance between the total export and import volumes. Contrarily to the PageRank-CheiRank description, it does not take into account the relative centrality of the country $c$ in the WTN. Otherwise stated, it does not take account of the possible strong indirect economical relations between countries. 

\begin{figure}[t!]
	\centering
	\includegraphics[width=0.95\columnwidth]{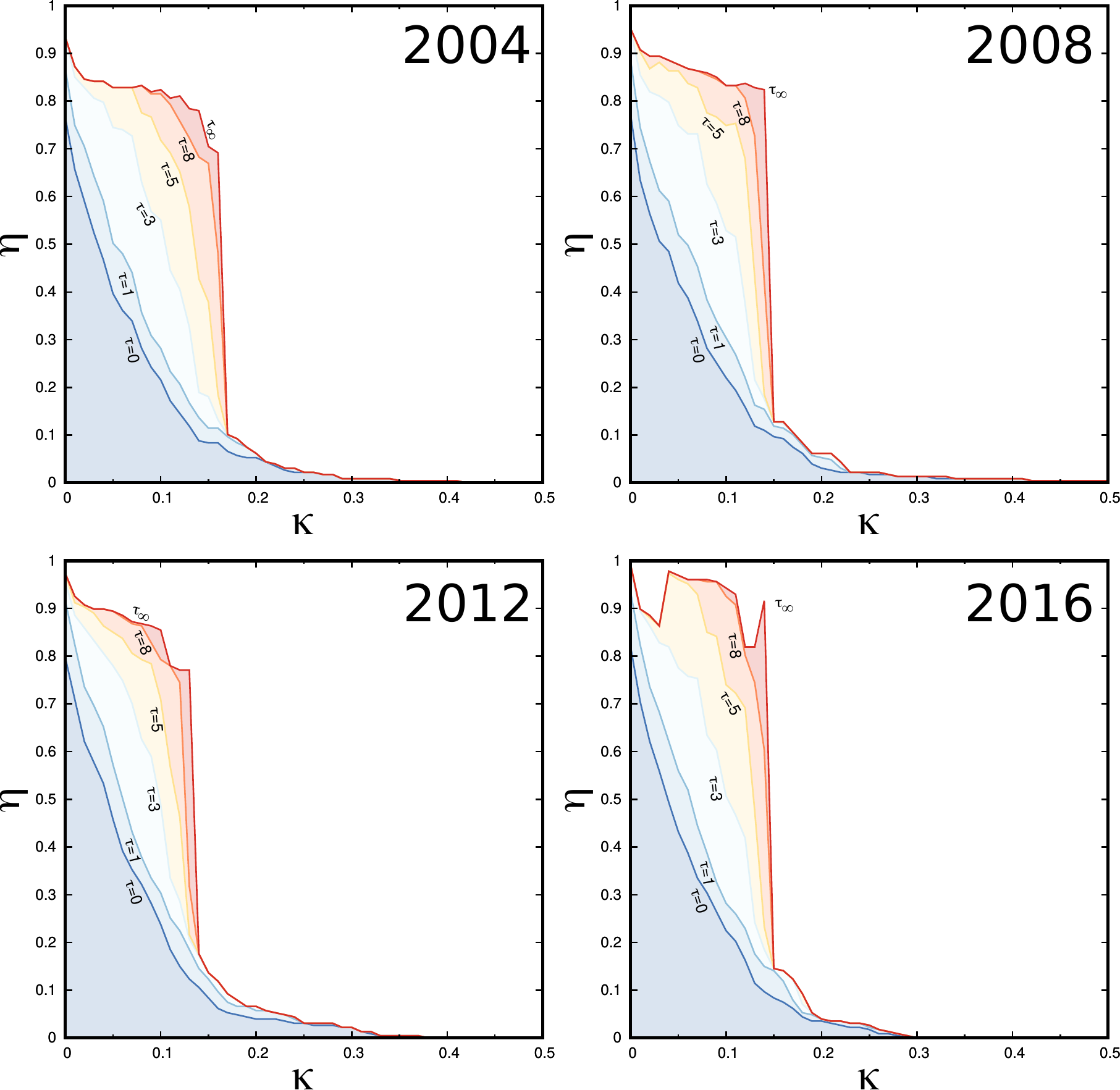}
	\caption{\csentence{Fraction of bankrupted countries for the WTN of 2004, 2008, 2012, and 2016.} Fraction $\eta$ of countries went to bankruptcy up to the $\tau$th stage of crisis contagion as a function of the bankruptcy threshold $\kappa$. The crisis contagion has been computed for the WTN of
		2004 (top left),
		2008 (top right),
		2012 (bottom left), and
		2016 (bottom right).
		Once a country goes to bankruptcy, it is prevented to import products with the exception of petroleum and gas (model A). The damping factor is $\alpha=0.5$.}
	\label{fig3}
\end{figure}

In the following, we analyze the crisis contagion in the WTN for different years using the PageRank-CheiRank trade balance $B_c$ with the model A and with $\alpha=0.5$. 
In Fig.~\ref{fig3}, for all the considered years, 2004, 2008, 2012, and 2016, we observe a similar phase transition from a regime of contained crisis contagion ($\kappa>\kappa_c$), for which the crisis only spreads over a small fraction (less than 10\%) of the countries, to a regime of global crisis contagion ($\kappa<\kappa_c$), for which the crisis spreads over about 90\% of the countries. For these years, the transition occurs at about the same critical bankruptcy threshold $\kappa_c\simeq0.14-0.175$. Some peculiarities are present for the 2016 WTN, for which, we observe an irregular and non monotonous profile in the $\left[0,\kappa_c\right]$ region. This is due to the interplay between the WTN rewiring occurring at the successive stages of the crisis contagion and the relative protection of the main petroleum and gas exporters since even countries which went to bankruptcy can import these commodities from these suppliers.
Such irregular profile is absent for the less realistic model B which exhibits an even more sharper phase transition than the model A (see Additional file 1 - Fig.~\ref{Sfig1}). As an example, for the 2016 WTN, with the model A, Russia never goes to bankruptcy in the bankruptcy threshold interval $\kappa<0.04$, it goes to bankruptcy in the $\kappa\in[0.04,0.11]$ interval, it stays safe for $\kappa\in[0.12,0.13]$, it goes to bankruptcy at $\kappa=0.14$, then it stays safe for $\kappa>0.14$. The intervals for which Russia goes to bankruptcy are concomitant with the bumps and the peaks observed for the 2016 WTN in the region $\kappa<\kappa_c$ (Fig.~\ref{fig3}). In the model A, the fall of Russia is responsible for a almost complete WTN crisis.

As seen in Fig.~\ref{fig3}, in the $\kappa\in[0,\kappa_c]$ region, about 90\% of the countries go to bankruptcy. The countries which remain safe at a bankruptcy threshold $\kappa=0.1$ (model A) are given in the Additional files 2 (Tab.~\ref{table:2004-safe-0_1}) for 2004, 3 (Tab.~\ref{table:2008-safe-0_1}) for 2008, 4 (Tab.~\ref{table:2012-safe-0_1}) for 20012, and 5 (Tab.~\ref{table:2016-safe-0_1}) for 2016. Most of these countries are petroleum and/or gas exporters, and, with the exception of 2016, for some of them petroleum and gas constitute the major volume of their exports, e.g., Nigeria (in 2004), Saudi Arabia (in 2004), Russia (in 2004, 2008, 2012), East Timor (in 2008). Also many of these remaining safe countries are islands, many of them being petroleum and gas exporters. We suppose that the other islands are peripheral in the WTN network and/or belong to some insulated minor trade exchange networks insensitive to the contagion. For the year 2016, for $\kappa=0.1$, the list of remaining safe countries is short, and even countries with a strong component of petroleum and gas in their export volume go to bankruptcy.

\begin{figure}[t!]
	\centering
	\includegraphics[width=0.95\columnwidth]{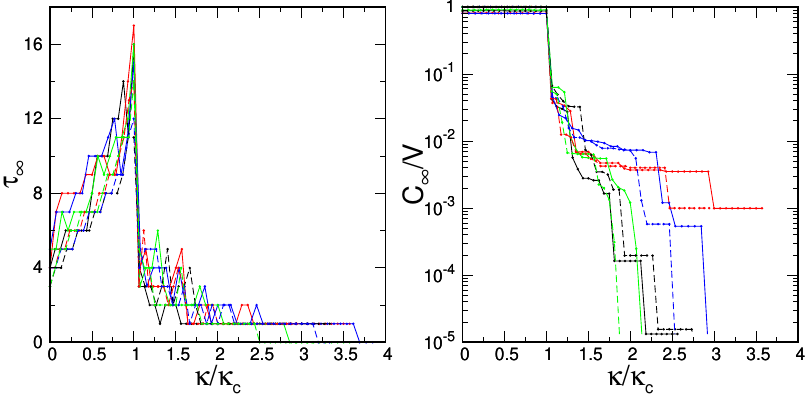}
	\caption{\csentence{(Left) Total number $\tau_{\infty}$ of crisis contagion stages as a function of the bankruptcy threshold $\kappa$ and (right) total crisis cost $C_\infty$ as a function of the bankruptcy threshold $\kappa$.}
	The total cost $C_\infty$ is defined according to the formula (\ref{eq:Cinf}), i.e., $C_\infty(\kappa)=C(\kappa,\tau_\infty)$.
	We use the WTN for years 2004 (black), 2008 (red), 2012 (blue), and 2016 (green).
		Solid lines correspond to the model A: once a country goes to bankruptcy, it is prevented to import products with the exception of petroleum and gas.
		Dashed lines correspond to the model B: once a country goes to bankruptcy, it is prevented to import any product. The lines allow to adapt an eye between the dots which represent the numerically computed values.
		The total amount of the World Trade transactions is
		$V=9.43\times10^{12}$~USD in 2004,
		$1.68\times10^{13}$~USD in 2008,
		$1.85\times10^{13}$~USD in 2012, and
		$1.62\times10^{13}$~USD in 2016.
		The damping factor is $\alpha=0.5$.}
	\label{fig4}
\end{figure}

The total number of crisis contagion stages, $\tau_\infty$, as a function of the bankruptcy threshold $\kappa$, also exhibits a phase transition (see Fig.~\ref{fig4} left) from a regime ($\kappa<\kappa_c$) for which $\tau_{\infty}$ rapidly increases with $\kappa$, from $\tau_\infty\simeq4$ for $\kappa=0$ up to $\tau_\infty\simeq12-16$ for $\kappa=\kappa_c$, and a regime ($\kappa>\kappa_c$) for which the crisis contagion stops after few stages, $\tau_{\infty}\lesssim5$. In the latter regime, it is not infrequent that the contagion even stops after $\tau_{\infty}=1$ or $2$ stages. We clearly observe that, for all the considered years, we obtain the same curve $\tau_\infty$ vs. $\kappa/\kappa_c$ whether we use the model A or the model B.

The phase transition is also clearly seen in the evolution of the total crisis cost $C_\infty$ (\ref{eq:Cinf}) as a function of the bankruptcy threshold $\kappa$ (Fig.~\ref{fig4} right). For $\kappa<\kappa_c$, the cost of the crisis is about 80-90\% of the total USD volume $V$ exchanged between the countries in the WTN. By contrast, for $\kappa>\kappa_c$, the total cost of the crisis is less than 5\% of $V$.
Such a graph could help any supranational agency to limit the cost of a crisis induced by the application of austerity policies to indebted countries. Indeed, the calculus of the PCBT (\ref{eq:PCTB}) allows to select a bankruptcy threshold limiting the crisis cost below a given value. Eg, for $\kappa\gtrsim1.5\kappa_c$, the cost of a crisis is less than the hundredth of the total volume exchanged. We also observe that the curves for all the considered years, whether we use the model A or the model B, fall into practically the same curve. Differences between different years are visible for $\kappa\gtrsim1.5\kappa_c$. In this region, going from $\kappa=3.5\kappa_c$ to $\kappa\simeq1.5\kappa_c$, the stairway structure of the curves is due to the successive sudden bankruptcies of countries at specific bankruptcy thresholds $\kappa$. These bankruptcies are dependent of the details of the WTN structure for the considered years. Let us also note that, in the region $\kappa<\kappa_c$, the total cost of the crisis is about 80-90\% of the total USD volume $V$ exchanged, the remaining 10-20\% of the volume $V$ still flows through the WTN since exports to the remaining 10\% of the countries are still allowed even for countries in bankruptcy.

In the following, we analyze with details the role of the countries in the crisis contagion.

\subsection*{Geographical distibution of the PageRank-CheiRank trade balance}

\begin{figure*}[t!]
	\centering
	\includegraphics[width=0.95\columnwidth]{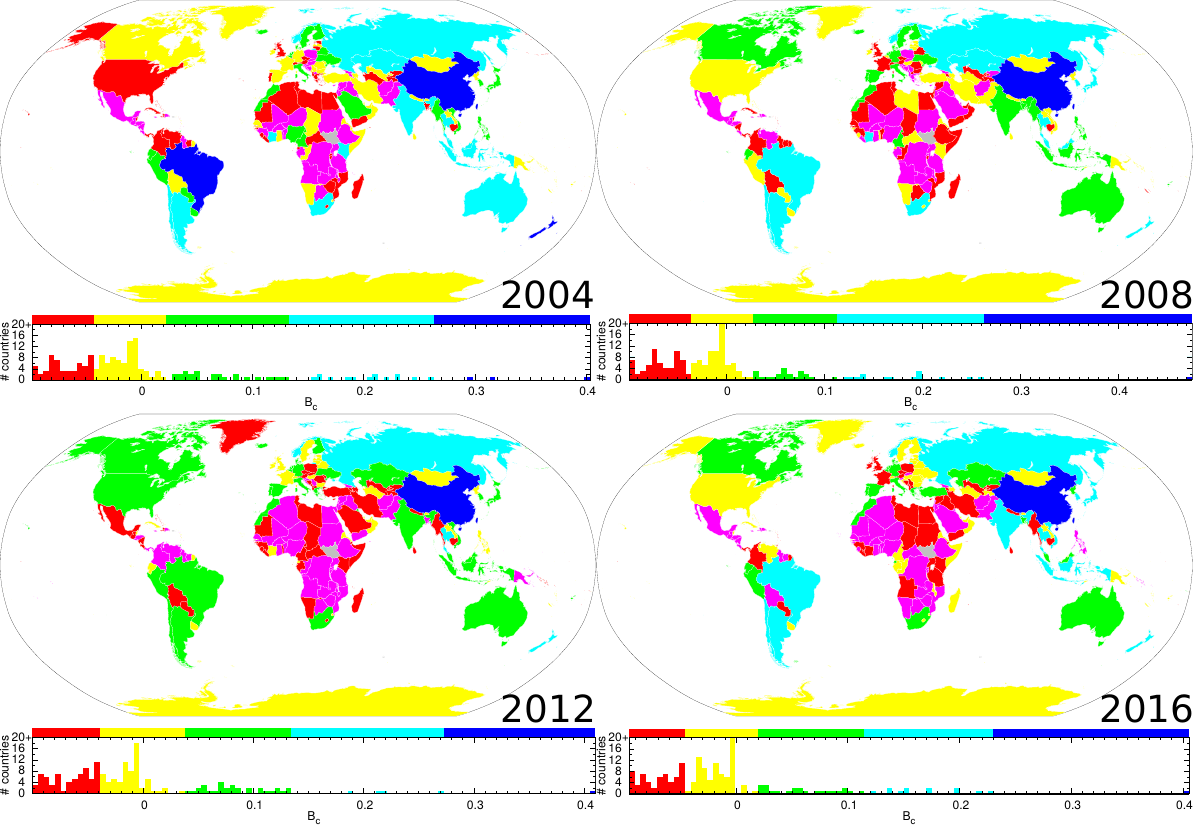}
	\caption{\csentence{Geographical distribution of the PageRank-CheiRank trade balance $B_{c}$ at the contagion stage $\tau = 0$ with the bankruptcy threshold $\kappa = 0.1$.} For each year, at the contagion stage $\tau = 0$, the countries colored in red (blue) have the most negative (positive) balance $B_c>-\kappa$.
	Color categories are obtained using the Jenks natural breaks classification method \cite{jenks}.
	Countries going to bankruptcy at contagion steps $\tau=0$ are colored in magenta. 
		The damping factor is $\alpha=0.5$.}
	\label{fig5}
\end{figure*}

In Fig.~\ref{fig5}, we present the PCTB for each country. As an example, let us consider that the bankruptcy threshold is $\kappa=0.1$. Hence, countries with $B_c<-\kappa=-0.1$ are the seeds of the crisis contagion. Among these countries, there are
many African countries including some Sub-Saharan countries, i.e., Mali, Niger, Burkina-Faso, DRC, Zambia (all considered years),
some Central American countries including Mexico (2004, 2008, 2016) and the Dominican Republic (all considered years), 
some Middle East countries including Israel (2004, 2012, 2016), Egypt (2012), Syria (2004, 2012), Irak (2004, 2008, 2012), and Saudi Arabia (2008),
some Asian countries including Afghanistan and Pakistan (all considered years except Pakistan for 2008), Papua New Guineas (2012), Bangladesh (2012), and Philippines (2016),
East Europe countries including Poland (2004, 2008), Slovakia (2004, 2008, 2016), the successor states of the former Yugoslavia (alternately during the considered years), Greece (2008), and Georgia (all considered years). More globally, countries with $B_c\lesssim0$ will certainly go to bankruptcy at the very first stage of the crisis contagion. These countries, with magenta, red, and yellow colors in Fig.~\ref{fig5}, are systematically the whole African continent excepting Morocco and South Africa, the Middle East, Laos, Cambodia, Papua New Guineas, the Central America and the Caribbean region, the northern South America, Bolivia, and Paraguay. Also, we note that North American countries, e.g., US (excepting for 2012), West European countries, e.g., France, UK, Ireland, Switzerland, and East European countries  have $B_c\lesssim0$.
Contrarily, the five BRICS appear to be among the most virtuous countries, with $0.1\lesssim B_c\lesssim0.4$.

\begin{figure*}[t!]
	\centering
	\includegraphics[width=0.95\columnwidth]{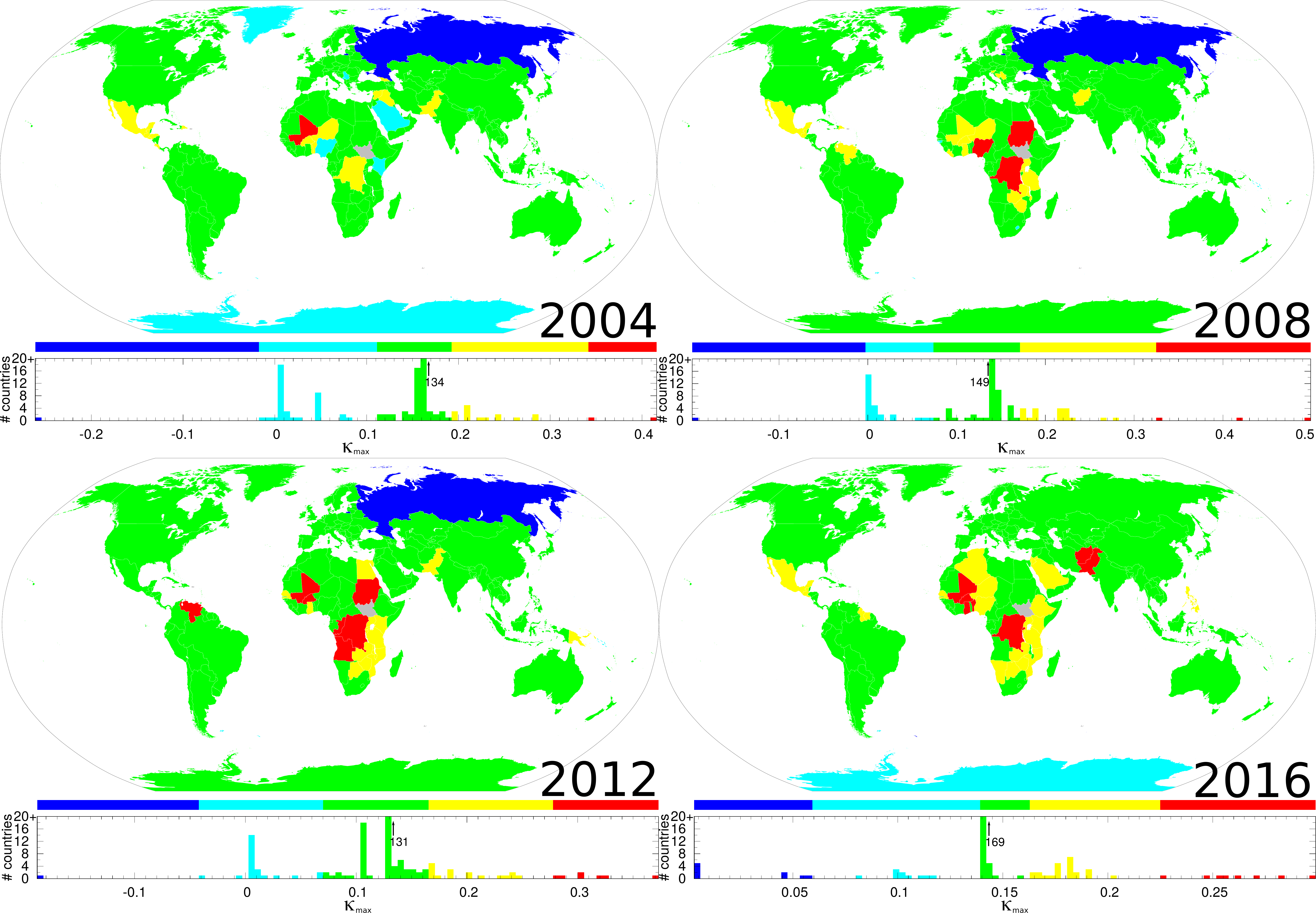}
	\caption{\csentence{Geographical distribution of the maximum bankruptcy threshold $\kappa_{\mbox{max}}$ at which a country goes to bankruptcy at any step of the contagion.}
		Countries with the highest (lowest) $\kappa_{\mbox{max}}$ are colored in red (blue).
		Color categories are obtained using the Jenks natural breaks classification method \cite{jenks}.	
		Here, once a country goes to bankruptcy, it is prevented to import products with the exception of petroleum and gas (model A).
		The damping factor is $\alpha=0.5$. At this scale, small sized islands are not visible. For information, the blue colored countries for the 2016 WTN are the following islands BV, IO, CC, HM, YT, AN, PN and GS (not visible in the world map).}
	\label{fig6}
\end{figure*}

Let us define for each country the maximum bankruptcy threshold $\kappa_{\mbox{max}}$ at which a country goes to bankruptcy at least at the final stage of the crisis contagion. Otherwise stated, for a $\kappa_{\mbox{max}}$ value associated to a given country $c$, this country do not go to bankruptcy for $\kappa>\kappa_{\mbox{max}}$, i.e., $B_c$ is always greater than $-\kappa$ at any stage of the contagion process for any $\kappa>\kappa_{\mbox{max}}$. Fig.~\ref{fig6} shows the geographical distribution of $\kappa_{\mbox{max}}$. For the model A, we observe that Russia is the less affected country by the crisis for the years 2004, 2008, and 2012 (for Russia, $\kappa_{\mbox{max}}\simeq-0.2$, i.e., at any crisis stage $\tau$, $B_c>-\kappa,\,\forall\kappa>-0.2$). Also, for 2004, Saudi Arabia, Nigeria, and Kenya are among the safest countries with $\kappa_{\mbox{max}}\lesssim0.1$ (i.e., these countries have $B_c>-\kappa,\; \forall\kappa\gtrsim0.1$). The fact that the four just above cited countries, for the above cited years, are the safest countries is due to their status of big petroleum and/or gas exporter. Russia has even $\kappa_{\mbox{max}}<0$, this means that, for the years 2004, 2008, and 2012, Russia occupies a peculiar protected position in the WTN.

For each considered years in Fig.~\ref{fig6}, we observe a peak in the country distribution at $\kappa_{\mbox{max}}$ just below $\kappa_c\simeq 0.175$ (2004), $0.15$ (2008), $0.14$ (2012), $0.15$ (2016). Such a country distribution can be used to precisely determine the critical bankruptcy threshold $\kappa_{c}$.

The most vulnerable countries (with $\kappa_{\mbox{max}}\gtrsim0.2$) are Central and South American countries (2004, 2008, 2016), including Mexico (2004, 2008, 2016), Guatemala (2004, 2008, 2016), El Salvador (2004, 2016), Honduras (2004), Costa Rica (2004), Dominican Republic (2004, 2008), Venezuela (2008, 2012), Guyana (2016), and Suriname (2016), Sub Saharan countries, including Mali (2004, 2008, 2012, 2016), Burkina Faso (2004, 2008, 2012, 2016),  Togo (2004), Benin (2004, 2016), Niger (2004, 2016), RDC (2004, 2008, 2012, 2016), Liberia (2008), Ghana (2008, 2012, 2016), Nigeria (2008), Sudan (2008, 2012), Uganda (2008, 2012, 2016), Rwanda (2008, 2012, 2016), Tanzania (2008, 2012, 2016), Zambia (2008, 2012, 2016), Zimbabwe (2008, 2012, 2016), Malawi (2008, 2012), Senegal (2012, 2016), Egypt (2012), Republic of Congo (2012), Angola (2012), Burundi (2012, 2016), Kenya (2012, 2016), Mozambique (2012, 2016), Bostwana (2012, 2016), Nigeria (2016), Ethiopia (2016), Algeria (2016), and Namibia (2016), Middle East countries, including Syria (2004), Iraq (2004), Georgia (2004), Egypt (2012), Israel (2016), Jordan (2016), and Saudi Arabia (2016), few European countries, Slovenia (2008), Bosnia-Herzegovina (2008) and Serbia (2008), and Asian countries, including Pakistan (2004, 2008, 2016), Afghanistan (2008, 2016), and Philippines (2016), and Papua New Guineas (2012). As a summary, for the considered year, the most fragile countries in the WTN are primarily many Sub Saharan countries, Central and South American countries, and some Middle East and Asian countries.

\begin{figure*}[t!]
	\centering
	\includegraphics[width=0.95\columnwidth]{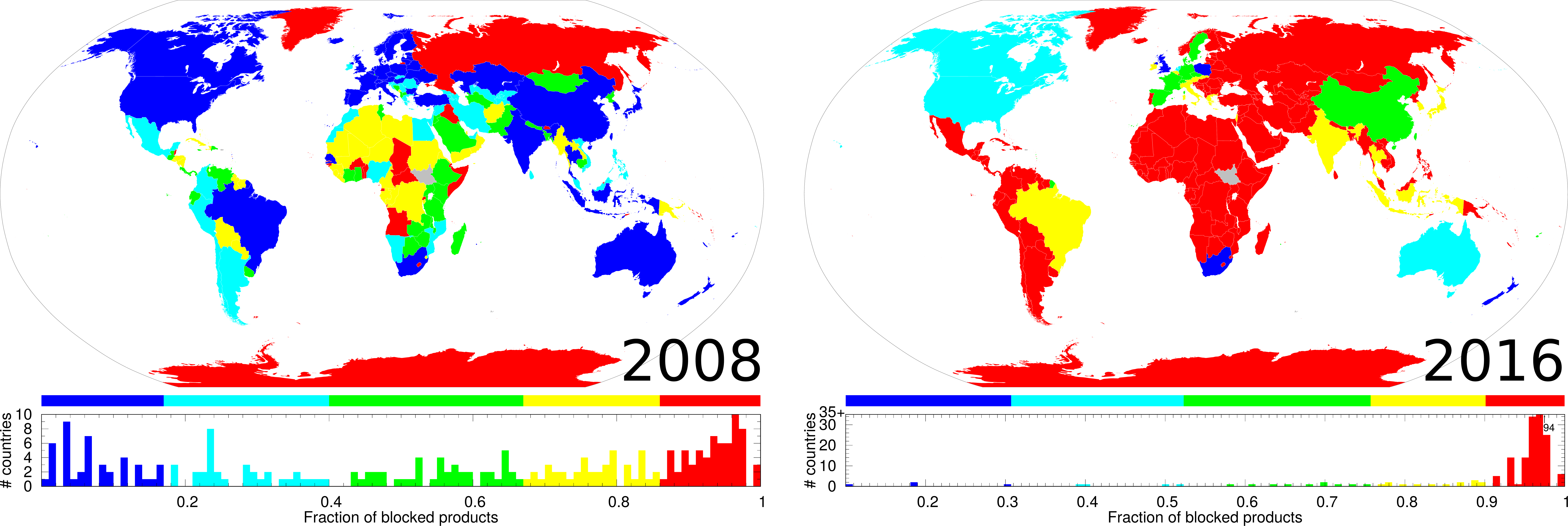}
	\caption{\csentence{Fraction of products which can not be exported by countries by lack of importers.} The color is function of the fraction of products which can not be exported by countries. Countries in blue can still export most of their products. Countries in red can almost no more export any of their products.
		Color categories are obtained using the Jenks natural breaks classification method \cite{jenks}.
		The computed data concern the 2008 and 2016 WTNs with $\kappa=0.1$ at $\tau_{\infty}$ and $\alpha = 0.5$. Once a country goes to bankruptcy, it is prevented to import products with the exception of petroleum and gas (model A).}
	\label{fig7}
\end{figure*}

\begin{figure*}[t!]
	\centering
	\includegraphics[width=0.7\columnwidth]{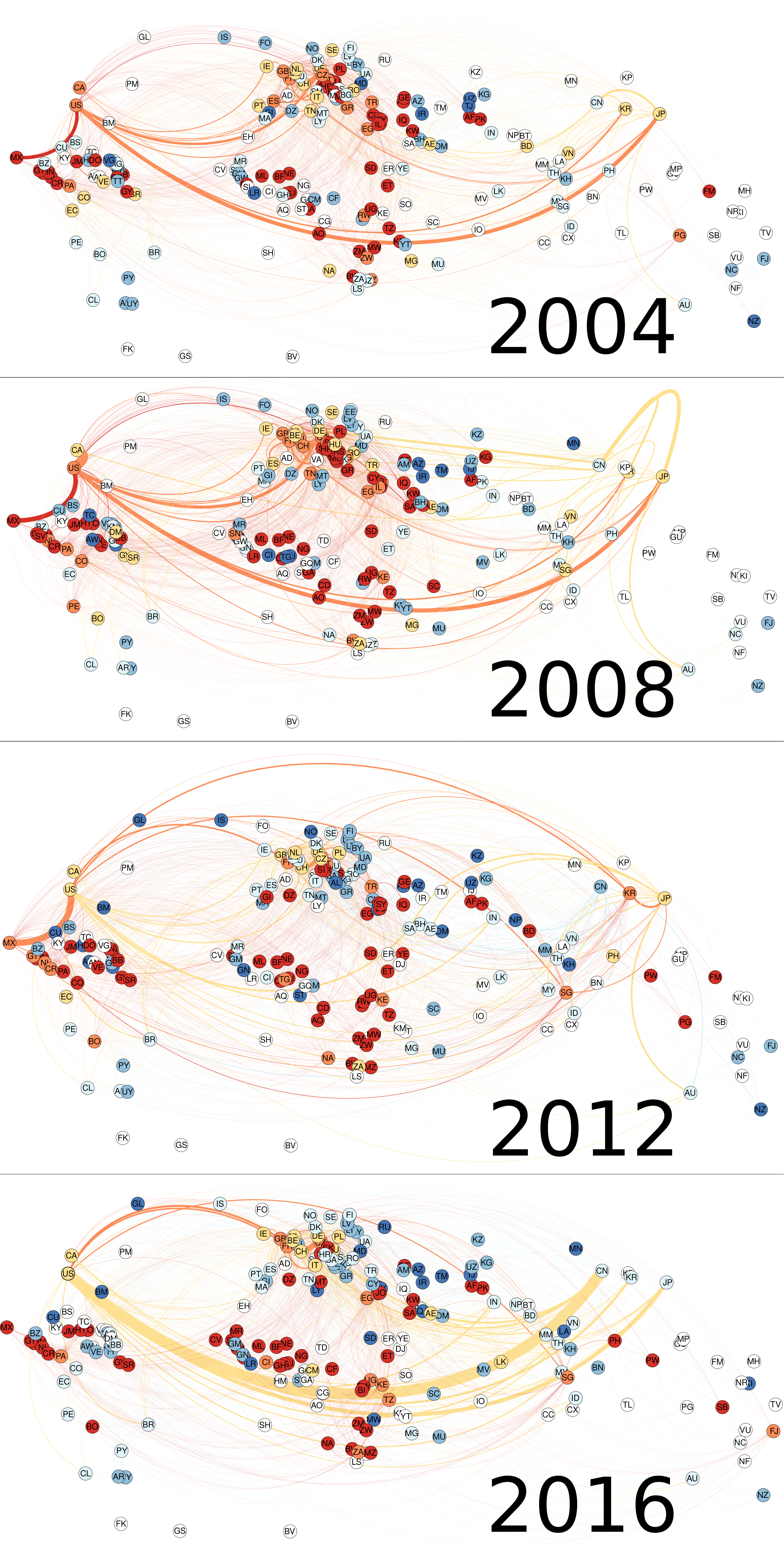}
	\caption{\csentence{Crisis contagion network for years 2004, 2008, 2012, and 2016.} The colors of country nodes range from red, for countries going to bankruptcy at stage $\tau=0$, to blue, for countries  going to bankruptcy at stage $\tau=\tau_{\infty}$. White nodes corresponds to countries which never go to bankruptcy.
		The direction of the link between two countries A and B is given by its curvature. If A points to B following the bent path in the counterclockwise direction (A$\smile$B) then A went to bankruptcy at the stage just before B, otherwise, i.e. (A$\frown$B), B went to bankruptcy at the stage just before A. The color of the link is the color of the node source. The width of the link is proportional to the export volume from the target country to the source country in the unmodified WTN . Here, once a country goes to bankruptcy, it is prevented to import products with the exception of petroleum and gas (model A). The bankruptcy threshold is $\kappa=0.1$ and the damping factor is $\alpha=0.5$.
	}
	\label{fig8}
\end{figure*}

The fact that bankrupted countries are prevented to import products implies that, during the contagion process, more and more products can not be exchanged. As an example, we show in Fig.~\ref{fig7}, for a bankruptcy threshold $\kappa=0.1$, the fraction of products which, at the end of the contagion, can not be exported by countries by lack of importers. For the 2008 WTN (Fig.~\ref{fig7}, left column), we observe that most of the countries of the Western world have less than 17\% of their exports blocked due to the crisis contagion. This means that at the end of the contagion process, these countries have at least one importer for almost each of their product. The same situation is found for some former USSR countries or satellites, such as Ukraine, Belarus, Moldova, Bulgaria, and Kazakhstan, some Middle Eastern countries, such as Turkey, and Asian countries, such as China, India, Thailand, Taiwan, South Korea, Japan, Singapore, and Indonesia. Although Russia do not go to bankruptcy during the crisis contagion at $\kappa=0.1$, nevertheless more than 87\% of its exports have been indirectly prevented by the crisis contagion. Russia remains safe in the 2008 WTN crisis contagion thanks to petroleum and gas exports which correspond to 60\% of the total Russian exports and which can be imported by any country in the model A. For the 2016 WTN (Fig.~\ref{fig7}, right column), the crisis is more severe as most of the countries have more than 90\% of their exports prevented. Only UK, Poland, South Africa, and New Zealand have less than 30\% of their exported products blocked.

\subsection*{Crisis contagion networks}

Let us define a network of causality where a country $c$ points to a country $c'$, if the country $c$ goes to bankruptcy at the crisis contagion stage $\tau$ and
the country $c'$ goes to bankruptcy at the next stage $\tau+1$. Otherwise stated, the bankruptcy of the country $c'$ follows right away the bankruptcy of the country $c$. In Fig.~\ref{fig8}, we show the network of crisis contagion causality for 2004, 2008, 2012, and 2016 WTNs and for a bankruptcy threshold $\kappa=0.1$. A country is colored according to the crisis contagion stage at which it goes to bankruptcy, from red for $\tau=0$ to blue for $\tau=\tau_\infty$. The direction of the links is given by the bending of the links, i.e., if the country $c$ points to the country $c'$ following a bent path in the counterclockwise direction, $c\smile c'$, then the country $c'$ goes to bankruptcy right after the country $c$. In the other hand, if the path direction from $c$ to $c'$ is clockwise, $c\frown c'$, then the country $c'$ goes to bankruptcy right before the country $c$. The width of the link from country $c$ to country $c'$ is proportional to the prevented export volume from country $c'$ to country $c$, i.e., $M_{cc'}=\sum_{p\in\tilde{\mathcal{P}}}M^p_{cc'}$.
The links are colored according to the color of the source. Consequently, the \textit{patients zero} of the crisis are the reddish countries and the very first banned trade exchanges are the reddish links. For all the considered years, the seeds of the crisis are mainly countries from Sub Sahara, Middle East, Central America, and Eastern Europe. We can observe only very few Asian countries as seeds of the crisis. The directions of the very first banned trade exchanges are meaningful. For the 2004 WTN, bunches of them come from Africa, Middle East, and Central America to Europe. Another bunches come from Eastern Europe and Central America to North America. Thus the fall of the US, which occurs at the second stage of the crisis contagion, stems mainly from the failures of Mexico, and Central American countries and East European countries. Once fallen, US drives to bankruptcy Western European countries and  ignite the crisis in Asia where Japan and South Korea go to bankruptcy at the third stage of the contagion. The failure of these latter countries then induce the failure of China and Australia. A similar contagion scheme occurs in the 2008 WTN. For the 2012 WTN, the US go to bankruptcy at the third stage of the crisis contagion after the failure of Mexico, South Korea, Singapore, and France. Singapore and South Korea propagate the crisis to Japan. For the 2016 WTN, the US also fall at the third contagion stage being impacted by the previous failure of Singapore, Great Britain, and France. Then, the failure of the US directly impacts China, South Korea, and Japan. An animation shows the contagion dynamics for the 2016 WTN (see \nameref{sec:add6}).

\begin{figure*}[t!]
	\centering
	\includegraphics[width=0.95\columnwidth]{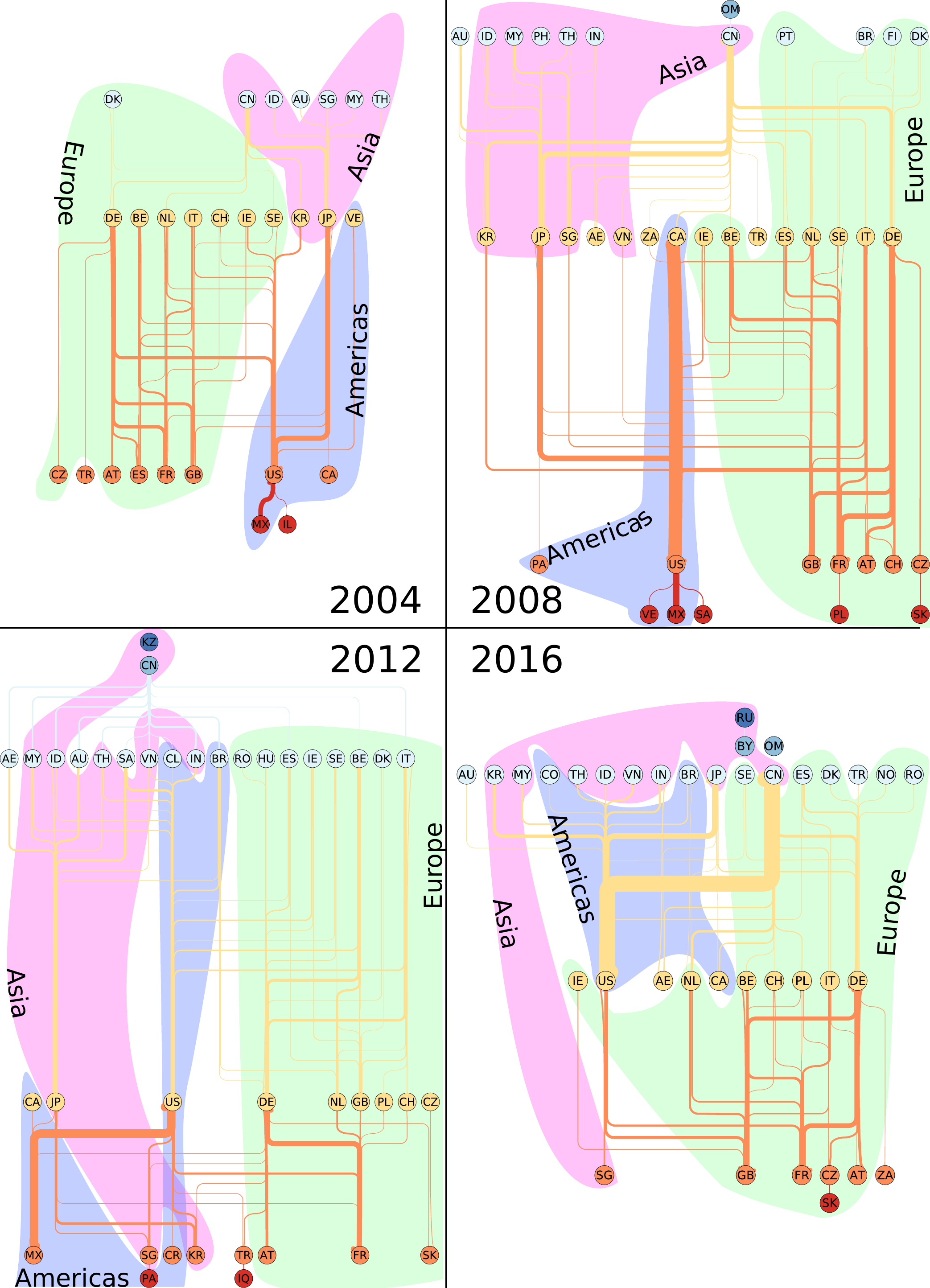}
	\caption{\csentence{Crisis contagion network with a hierarchical layout for years 2004, 2008, 2012, and 2016.} Only imports greater than $10^{10}$ USD are represented. Countries going to bankruptcy at the same contagion step $\tau$ are aligned in the same row. From bottom (red country nodes) to top (blue country nodes), the rows are associated to contagion step from $\tau=0$ to $\tau=3$ for 2004, $\tau=4$ for 2008, $\tau=5$ for 2012 and 2016. The width of the link going from a country $c$ which goes to bankruptcy at the stage $\tau$ to a country $c'$ which goes to bankruptcy at the stage $\tau'=\tau+1$ is proportional to the volume usually imported by country $c$ from $c'$.
		Here, once a country goes to bankruptcy, it is prevented to import products with the exception of petroleum and gas (model A).
		Colored zones gather countries from the same continent (green for European countries, blue for American countries, and pink for Asian countries).
		The bankruptcy threshold is $\kappa=0.1$ and the damping factor is $\alpha=0.5$.}
	\label{fig9}
\end{figure*}

Let us focus on the greatest volume trade exchanges between countries. Fig.~\ref{fig9} shows the hierarchy of the crisis contagion causality for imports greater than $10^{10}$ USD (as a complement, Fig.~\ref{Sfig2} shows the same data but geographically distributed).

For the 2004 WTN, among the big exporters, Mexico and Israel contribute to the fall of the US. From the fall of the US, one of the main paths of contagion can be followed, US $\rightarrow$ JP $\rightarrow$ (Asian countries and Australia). The bankruptcy of all the European countries are due to the conjugated effect of the fall at stage $\tau=1$ of the US and of part of the main European economies. Bankruptcies of the European countries, and also of South Korea, contribute then to the fall of the Asian big exporters (such as China) and of Australia, which, as stressed before, are the last countries to go to bankruptcy. Let us note that France and Great Britain also contribute to the failure of Japan.

For the 2008 WTN, we obtain a similar scenario excepting that Venezuela and Saudi Arabia, in addition to Mexico, lead the US to the bankruptcy. Also, Eastern Europe countries, Poland and Slovakia, are the seeds of the contagion in Europe. We can observe that the BRIC, i.e., Brazil, India, and China, are among the countries which are the last affected by the contagion.
This remark is also true for 2012 and 2016 WTNs. Russia, which is a petroleum and gas exporter, is never affected by the contagion excepting for the 2016 WTN.

For the 2012 WTN, at the crisis stage $\tau=0$ and $\tau=1$, there are several sources of contagion: Central America with Mexico, Panama and Costa Rica, Europe with Austria, France, and Slovakia, Middle East with Iraq and Turkey, and Asia with Singapore and South Korea. The crisis is already  present, at the very first stages of the contagion, in all the continents excepting Oceania. At stage $\tau=2$, the US are mainly affected by the previous fall of some Central American countries, some Asian countries and France. Then the US contributes to propagate the crisis to the rest of the world. The crisis in Asia is also brought by the fall of Japan induced by the bankruptcy of Singapore and South Korea, and in a somewhat lesser importance by the bankruptcy of France. The bankruptcy of the rest of European countries follows mainly the fall of France, and Austria, and follows secondarily the fall of Turkey, South Korea, Singapore, and Mexico.

For the 2016 WTN, the European countries ignite the crisis with Slovakia as seed of the contagion. The crisis propagates to North American countries and then to Asia and the rest of the world. We note that in addition to European countries Singapore is also in bankruptcy at the early stages of the contagion and contributes to the fall of the US. Here, Russia is the last country going to bankruptcy.

\section*{Conclusion and discussion}

The Google matrix analysis of the world trade network allows to probe the direct and indirect trade exchange dependencies between countries. Unlike the simple \textit{accounting} view obtained from the usual import-export balance, relying on the total volumes of exchanged commodities between countries (\ref{eq:IEB}), the PageRank-CheiRank trade balance (PCTB) (\ref{eq:PCTB}) allows to take account of the long range inter-dependencies between world economies. The WTN crisis contagion model is build upon the iterative measure of the PCTB for each country. Once a country have a PCTB below a threshold $-\kappa$, it is declared in a bankruptcy state in which it can no more import commodities excepting some vital one for the industry, i.e., petroleum and gas. This state corresponds either to the fact that a country with a very negative trade balance have not enough liquidity to import non essential commodities, or to the decision of a supranational economic authority trying to contain a crisis by placing an unhealthy national economy in bankruptcy. The bankruptcies of economies with PCTB less than $-\kappa$ induce a rewiring of the world trade network which possibly weaken other economies. In the phase corresponding to a bankruptcy threshold $\kappa>\kappa_c$, the crisis contagion is rapid and contained since it affects only less than 10\% of the world countries and induces a total cost of less than 5\% of the total USD volume exchanged in the WTN. This total cost of the crisis drops exponentially with the increase of $\kappa$. In the phase corresponding to a bankruptcy threshold $\kappa<\kappa_c$, the cascade of bankruptcies can not be contained and the crisis is global, affecting about 90\% of the world countries. The bankruptcy threshold $\kappa$ is the order parameter of the phase transition. In the global crisis phase ($\kappa<\kappa_c$), at the first stage ($\tau=0$) of the contagion, myriads of countries with low exchanged volume (ie, low import and export volumes) go to bankruptcy. These countries belong mainly to Sub Saharan Africa, Central and South America, Middle East, and Eastern Europe. In the next stage of the crisis contagion, the conjugated effect of the bankruptcies of these countries contribute to the fall of big exporters, such as the US or Western European countries. As an example, for 2004, 2012, and 2016 WTNs, the bankruptcy of France at the contagion stage $\tau=1$ is solely due to the failure of many low exchanged volume countries, which, here, individually import from France a volume of commodities less than $10^{10}$ USD. Otherwise stated, France failure is caused by the failure of many small importers. Great Britain is a similar case for the 2004, 2008, and 2016 WTNs. Among the big exporters (ie, with a exchanged volume greater than $10^{10}$ USD), European and American countries are the sources of the crisis contagion. The gates from which crisis enters Asia are usually Japan, Korea, and Singapore. Generally, Asian countries go to bankruptcy at the end of the crisis contagion, with China, India, Indonesia, Malaysia and Thailand, being, with Australia, usually the last economies to fall. We also observe that failures of the four BRIC occur during the last stages of the crisis contagion.

As a future development of the presented WTN crisis contagion analysis, it would be interesting to study the cascades of country bankruptcies induced by a sharp increase of the price of a given commodity. Indeed, within our model, such an increase of the price of petroleum and/or gas would highlight the structural vulnerability of the countries to an energy crisis contagion.

\begin{backmatter}

\section*{Availability of data and materials}

The raw data is available from the UN Comtrade database \cite{comtrade}. Additional output data and/or plots of data generated are available upon request.

\section*{Abbreviations}
\label{sec:abbreviation}
WTN: World trade network;
PCTB: PageRank-CheiRank trade balance;
DRC: Democratic Republic of the Congo;
BRIC: Brazil, Russia, India, China;
BRICS: Brazil, Russia, India, China, South Africa.

ISO 3166-1 alpha-2 code for countries:

AF: Afghanistan; AL: Albania; DZ: Algeria; AS: American Samoa; AD: Andorra; AO: Angola; AI: Anguilla; AQ: Antarctica; AG: Antigua and Barbuda; AR: Argentina; AM: Armenia; AW: Aruba; AU: Australia; AT: Austria; AZ: Azerbaijan; BS: The Bahamas; BH: Bahrain; BD: Bangladesh; BB: Barbados; BY: Belarus; BE: Belgium; BZ: Belize; BJ: Benin; BM: Bermuda; BT: Bhutan; BO: Bolivia; BA: Bosnia and Herzegovina; BW: Botswana; BV: Bouvet Island; IO: British Indian Ocean Territory; VG: British Virgin Islands; BR: Brazil; BN: Brunei; BG: Bulgaria; BF: Burkina Faso; BI: Burundi; KH: Cambodia; CM: Cameroon; CA: Canada; CV: Cape Verde; KY: Cayman Islands; CF: Central African Republic; TD: Chad; CL: Chile; CN: China; CX: Christmas Island; CC: Cocos (Keeling) Islands; CO: Colombia; KM: Comoros; CG: Republic of the Congo; CK: Cook Islands; CR: Costa Rica; CI: Ivory Coast; HR: Croatia; CU: Cuba; CY: Cyprus; CZ: Czech Republic; KP: North Korea; CD: Democratic Republic of the Congo; DK: Denmark; DJ: Djibouti; DM: Dominica; DO: Dominican Republic; EC: Ecuador; EG: Egypt; SV: El Salvador; GQ: Equatorial Guinea; ER: Eritrea; EE: Estonia; ET: Ethiopia; FO: Faroe Islands; FK: Falkland Islands; FJ: Fiji; FI: Finland; FR: France; PF: French Polynesia; FM: Micronesia; GA: Gabon; GM: The Gambia; GE: Georgia; DE: Germany; GH: Ghana; GI: Gibraltar; GR: Greece; GL: Greenland; GD: Grenada; GU: Guam; GT: Guatemala; GN: Guinea; GW: Guinea-Bissau; GY: Guyana; HT: Haiti; HM: Heard Island and McDonald Islands; VA: Vatican; HN: Honduras; HU: Hungary; IS: Iceland; IN: India; ID: Indonesia; IR: Iran; IQ: Iraq; IE: Ireland; IL: Israel; IT: Italy; JM: Jamaica; JP: Japan Ryukyu Island; JO: Jordan; KZ: Kazakhstan; KE: Kenya; KI: Kiribati; KW: Kuwait; KG: Kyrgyzstan; LA: Laos; LV: Latvia; LB: Lebanon; LS: Lesotho; LR: Liberia; LY: Libya; LT: Lithuania; LU: Luxembourg; MG: Madagascar; MW: Malawi; MY: Malaysia; MV: Maldives; ML: Mali; MT: Malta; MH: Marshall Islands; MR: Mauritania; MU: Mauritius; YT: Mayotte; MX: Mexico; MN: Mongolia; ME: Montenegro; MS: Montserrat; MA: Morocco; MZ: Mozambique; MM: Myanmar; MP: Northern Mariana Islands; NA: Namibia; NR: Nauru; NP: Nepal; AN: Netherlands Antilles; NL: Netherlands; NC: New Caledonia; NZ: New Zealand; NI: Nicaragua; NE: Niger; NG: Nigeria; NU: Niue; NF: Norfolk Islands; NO: Norway; PS: State of Palestine; OM: Oman; PK: Pakistan; PW: Palau; PA: Panama; PG: Papua New Guinea; PY: Paraguay; PE: Peru; PH: Philippines; PN: Pitcairn; PL: Poland; PT: Portugal; QA: Qatar; KR: South Korea; MD: Moldova; RO: Romania; RU: Russia; RW: Rwanda; SH: Saint Helena; KN: Saint Kitts and Nevis; LC: Saint Lucia; PM: Saint Pierre and Miquelon; VC: Saint Vincent and the Grenadines; WS: Samoa; SM: San Marino; ST: Sao Tome and Principe; SA: Saudi Arabia; SN: Senegal; RS: Serbia; SC: Seychelles; SL: Sierra Leone; SG: Singapore; SK: Slovakia; SI: Slovenia; SB: Solomon Islands; SO: Somalia; ZA: South Africa; GS: South Georgia and the South Sandwich Islands; ES: Spain; LK: Sri Lanka; SD: Sudan; SR: Suriname; SZ: Swaziland; SE: Sweden; CH: Switzerland; SY: Syria; TJ: Tajikistan; MK: Macedonia; TH: Thailand; TL: Timor-Leste; TG: Togo; TK: Tokelau; TO: Tonga; TT: Trinidad and Tobago; TN: Tunisia; TR: Turkey; TM: Turkmenistan; TC: Turks and Caicos Islands; TV: Tuvalu; UG: Uganda; UA: Ukraine; AE: United Arab Emirates; GB: United Kingdom; TZ: Tanzania; UM: United States Minor Outlying Islands; UY: Uruguay; US: United States; UZ: Uzbekistan; VU: Vanuatu; VE: Venezuela; VN: Vietnam; WF: Wallis and Futuna; EH: Western Sahara; YE: Yemen; ZM: Zambia; ZW: Zimbabwe.





\newcommand{\BMCxmlcomment}[1]{}

\BMCxmlcomment{

<refgrp>

<bibl id="B1">
  <title><p>Contagion in financial networks</p></title>
  <aug>
    <au><snm>Gai</snm><fnm>P</fnm></au>
    <au><snm>Kapadia</snm><fnm>S</fnm></au>
  </aug>
  <source>Proceedings of the Royal Society A: Mathematical, Physical and
  Engineering Sciences</source>
  <pubdate>2010</pubdate>
  <volume>466</volume>
  <issue>2120</issue>
  <fpage>2401</fpage>
  <lpage>2423</lpage>
  <url>https://royalsocietypublishing.org/doi/abs/10.1098/rspa.2009.0410</url>
</bibl>

<bibl id="B2">
  <title><p>Financial Networks and Contagion</p></title>
  <aug>
    <au><snm>Elliott</snm><fnm>M</fnm></au>
    <au><snm>Golub</snm><fnm>B</fnm></au>
    <au><snm>Jackson</snm><fnm>MO</fnm></au>
  </aug>
  <source>American Economic Review</source>
  <pubdate>2014</pubdate>
  <volume>104</volume>
  <issue>10</issue>
  <fpage>3115</fpage>
  <lpage>53</lpage>
</bibl>

<bibl id="B3">
  <title><p>The credit quality channel: Modeling contagion in the interbank
  market</p></title>
  <aug>
    <au><snm>Fink</snm><fnm>K</fnm></au>
    <au><snm>Krüger</snm><fnm>U</fnm></au>
    <au><snm>Meller</snm><fnm>B</fnm></au>
    <au><snm>Wong</snm><fnm>LH</fnm></au>
  </aug>
  <source>Journal of Financial Stability</source>
  <pubdate>2016</pubdate>
  <volume>25</volume>
  <fpage>83</fpage>
  <lpage>97</lpage>
  <url>http://www.sciencedirect.com/science/article/pii/S1572308916300444</url>
</bibl>

<bibl id="B4">
  <title><p>Energy crisis --- {Wikipedia}{,} The Free Encyclopedia</p></title>
  <aug>
    <au><cnm>{Wikipedia contributors}</cnm></au>
  </aug>
  <source>\url{https://en.wikipedia.org/w/index.php?title=Energy_crisis&oldid=928811598}</source>
  <pubdate>2019</pubdate>
  <note>[Online; accessed 2-February-2020]</note>
</bibl>

<bibl id="B5">
  <title><p>Oil crisis --- {Encyclop\ae dia Britannica}</p></title>
  <aug>
    <au><cnm>{Kettell, S.}</cnm></au>
  </aug>
  <source>\url{https://www.britannica.com/topic/oil-crisis}</source>
  <pubdate>2020</pubdate>
  <note>[Online; accessed 2-February-2020]</note>
</bibl>

<bibl id="B6">
  <title><p>United Nations Commodity Trade Statistics Database</p></title>
  <url>http://comtrade.un.org/db/</url>
</bibl>

<bibl id="B7">
  <title><p>The anatomy of a large-scale hypertextual Web search
  engine</p></title>
  <aug>
    <au><snm>Brin</snm><fnm>S</fnm></au>
    <au><snm>Page</snm><fnm>L</fnm></au>
  </aug>
  <source>Computer Networks and ISDN Systems</source>
  <pubdate>1998</pubdate>
  <volume>30</volume>
  <issue>1</issue>
  <fpage>107</fpage>
  <lpage>117</lpage>
  <url>http://www.sciencedirect.com/science/article/pii/S016975529800110X</url>
  <note>Proceedings of the Seventh International World Wide Web
  Conference</note>
</bibl>

<bibl id="B8">
  <title><p>Google's PageRank and Beyond: The Science of Search Engine
  Rankings</p></title>
  <aug>
    <au><snm>Langville</snm><fnm>AN</fnm></au>
    <au><snm>Meyer</snm><fnm>CD</fnm></au>
  </aug>
  <publisher>USA: Princeton University Press</publisher>
  <pubdate>2012</pubdate>
</bibl>

<bibl id="B9">
  <title><p>Google matrix analysis of directed networks</p></title>
  <aug>
    <au><snm>Ermann</snm><fnm>L.</fnm></au>
    <au><snm>Frahm</snm><fnm>K.M.</fnm></au>
    <au><snm>Shepelyansky</snm><fnm>D.L.</fnm></au>
  </aug>
  <source>Rev. Mod. Phys.</source>
  <publisher>American Physical Society</publisher>
  <pubdate>2015</pubdate>
  <volume>87</volume>
  <fpage>1261</fpage>
  <lpage>-1310</lpage>
  <url>https://link.aps.org/doi/10.1103/RevModPhys.87.1261</url>
</bibl>

<bibl id="B10">
  <title><p>Google Matrix of the World Trade Network</p></title>
  <aug>
    <au><snm>Ermann</snm><fnm>L.</fnm></au>
    <au><snm>Shepelyansky</snm><fnm>D.L.</fnm></au>
  </aug>
  <pubdate>2011</pubdate>
  <volume>120</volume>
  <fpage>A158</fpage>
  <lpage>-A171</lpage>
</bibl>

<bibl id="B11">
  <title><p>Google matrix analysis of the multiproduct world trade
  network</p></title>
  <aug>
    <au><snm>Ermann</snm><fnm>L</fnm></au>
    <au><snm>Shepelyansky</snm><fnm>DL</fnm></au>
  </aug>
  <source>The European Physical Journal B</source>
  <pubdate>2015</pubdate>
  <volume>88</volume>
  <issue>4</issue>
  <fpage>84</fpage>
  <url>https://doi.org/10.1140/epjb/e2015-60047-0</url>
</bibl>

<bibl id="B12">
  <title><p>Google matrix of Bitcoin network</p></title>
  <aug>
    <au><cnm>{Ermann, L.}</cnm></au>
    <au><cnm>{Frahm, K.M.}</cnm></au>
    <au><cnm>{Shepelyansky, D.L.}</cnm></au>
  </aug>
  <source>Eur. Phys. J. B</source>
  <pubdate>2018</pubdate>
  <volume>91</volume>
  <issue>6</issue>
  <fpage>127</fpage>
  <url>https://doi.org/10.1140/epjb/e2018-80674-y</url>
</bibl>

<bibl id="B13">
  <title><p>Contagion in Bitcoin Networks</p></title>
  <aug>
    <au><snm>Coquid{\'e}</snm><fnm>C</fnm></au>
    <au><snm>Lages</snm><fnm>J</fnm></au>
    <au><snm>Shepelyansky</snm><fnm>DL</fnm></au>
  </aug>
  <source>Business Information Systems Workshops</source>
  <publisher>Cham: Springer International Publishing</publisher>
  <editor>Abramowicz, Witold and Corchuelo, Rafael</editor>
  <pubdate>2019</pubdate>
  <fpage>208</fpage>
  <lpage>-219</lpage>
</bibl>

<bibl id="B14">
  <title><p>Patterns of dominant flows in the world trade web</p></title>
  <aug>
    <au><snm>Serrano</snm><fnm>M{\'A}</fnm></au>
    <au><snm>Bogu{\~n}{\'a}</snm><fnm>M</fnm></au>
    <au><snm>Vespignani</snm><fnm>A</fnm></au>
  </aug>
  <source>Journal of Economic Interaction and Coordination</source>
  <pubdate>2007</pubdate>
  <volume>2</volume>
  <issue>2</issue>
  <fpage>111</fpage>
  <lpage>124</lpage>
  <url>https://doi.org/10.1007/s11403-007-0026-y</url>
</bibl>

<bibl id="B15">
  <title><p>World-trade web: Topological properties, dynamics, and
  evolution</p></title>
  <aug>
    <au><snm>Fagiolo</snm><fnm>G</fnm></au>
    <au><snm>Reyes</snm><fnm>J</fnm></au>
    <au><snm>Schiavo</snm><fnm>S</fnm></au>
  </aug>
  <source>Phys. Rev. E</source>
  <publisher>American Physical Society</publisher>
  <pubdate>2009</pubdate>
  <volume>79</volume>
  <fpage>036115</fpage>
  <url>https://link.aps.org/doi/10.1103/PhysRevE.79.036115</url>
</bibl>

<bibl id="B16">
  <title><p>Structure and Response in the World Trade Network</p></title>
  <aug>
    <au><snm>He</snm><fnm>J</fnm></au>
    <au><snm>Deem</snm><fnm>MW</fnm></au>
  </aug>
  <source>Phys. Rev. Lett.</source>
  <publisher>American Physical Society</publisher>
  <pubdate>2010</pubdate>
  <volume>105</volume>
  <fpage>198701</fpage>
  <url>https://link.aps.org/doi/10.1103/PhysRevLett.105.198701</url>
</bibl>

<bibl id="B17">
  <title><p>The evolution of the world trade web: a weighted-network
  analysis</p></title>
  <aug>
    <au><snm>Fagiolo</snm><fnm>G</fnm></au>
    <au><snm>Reyes</snm><fnm>J</fnm></au>
    <au><snm>Schiavo</snm><fnm>S</fnm></au>
  </aug>
  <source>Journal of Evolutionary Economics</source>
  <pubdate>2010</pubdate>
  <volume>20</volume>
  <fpage>479</fpage>
  <lpage>514</lpage>
  <url>https://link.aps.org/doi/10.1007/s00191-009-0160-x</url>
</bibl>

<bibl id="B18">
  <title><p>Multinetwork of international trade: A commodity-specific
  analysis</p></title>
  <aug>
    <au><snm>Barigozzi</snm><fnm>M</fnm></au>
    <au><snm>Fagiolo</snm><fnm>G</fnm></au>
    <au><snm>Garlaschelli</snm><fnm>D</fnm></au>
  </aug>
  <source>Phys. Rev. E</source>
  <publisher>American Physical Society</publisher>
  <pubdate>2010</pubdate>
  <volume>81</volume>
  <fpage>046104</fpage>
  <url>https://link.aps.org/doi/10.1103/PhysRevE.81.046104</url>
</bibl>

<bibl id="B19">
  <title><p>The World Trade Network</p></title>
  <aug>
    <au><snm>De Benedictis</snm><fnm>L</fnm></au>
    <au><snm>Tajoli</snm><fnm>L</fnm></au>
  </aug>
  <source>The World Economy</source>
  <pubdate>2011</pubdate>
  <volume>34</volume>
  <issue>8</issue>
  <fpage>1417</fpage>
  <lpage>1454</lpage>
  <url>https://onlinelibrary.wiley.com/doi/abs/10.1111/j.1467-9701.2011.01360.x</url>
</bibl>

<bibl id="B20">
  <title><p>Hubs and Authorities in the World Trade Network Using a Weighted
  HITS Algorithm</p></title>
  <aug>
    <au><snm>Deguchi</snm><fnm>T</fnm></au>
    <au><snm>Takahashi</snm><fnm>K</fnm></au>
    <au><snm>Takayasu</snm><fnm>H</fnm></au>
    <au><snm>Takayasu</snm><fnm>M</fnm></au>
  </aug>
  <source>PLOS ONE</source>
  <publisher>Public Library of Science</publisher>
  <pubdate>2014</pubdate>
  <volume>9</volume>
  <issue>7</issue>
  <fpage>1</fpage>
  <lpage>16</lpage>
  <url>https://doi.org/10.1371/journal.pone.0100338</url>
</bibl>

<bibl id="B21">
  <title><p>Interconnected Banks and Systemically Important
  Exposures</p></title>
  <aug>
    <au><snm>Roncoroni</snm><fnm>A</fnm></au>
    <au><snm>Battiston</snm><fnm>S</fnm></au>
    <au><snm>D'Errico</snm><fnm>M</fnm></au>
    <au><snm>Ha{\l}aj</snm><fnm>G</fnm></au>
    <au><snm>Kok</snm><fnm>C</fnm></au>
  </aug>
  <source>SSRN</source>
  <publisher>Public Library of Science</publisher>
  <pubdate>2019</pubdate>
  <url>https://ssrn.com/abstract=3491235</url>
  <note>ECB Working Paper No. 2331,
  \url{https://ssrn.com/abstract=3491235}</note>
</bibl>

<bibl id="B22">
  <title><p>Influence of petroleum and gas trade on EU economies from the
  reduced Google matrix analysis of UN COMTRADE data</p></title>
  <aug>
    <au><snm>Coquid{\'e}</snm><fnm>C</fnm></au>
    <au><snm>Ermann</snm><fnm>L</fnm></au>
    <au><snm>Lages</snm><fnm>J</fnm></au>
    <au><snm>Shepelyansky</snm><fnm>DL</fnm></au>
  </aug>
  <source>The European Physical Journal B</source>
  <pubdate>2019</pubdate>
  <volume>92</volume>
  <issue>8</issue>
  <fpage>171</fpage>
  <url>https://doi.org/10.1140/epjb/e2019-100132-6</url>
</bibl>

<bibl id="B23">
  <title><p>Interdependence of sectors of economic activities for world
  countries from the reduced Google matrix analysis of WTO data</p></title>
  <aug>
    <au><snm>Coquidé</snm><fnm>C</fnm></au>
    <au><snm>Lages</snm><fnm>J</fnm></au>
    <au><snm>Shepelyansky</snm><fnm>DL</fnm></au>
  </aug>
  <pubdate>2019</pubdate>
</bibl>

<bibl id="B24">
  <title><p>Jenks natural breaks optimization --- {Wikipedia}{,} The Free
  Encyclopedia</p></title>
  <aug>
    <au><cnm>{Wikipedia contributors}</cnm></au>
  </aug>
  <source>\url{https://en.wikipedia.org/w/index.php?title=Jenks_natural_breaks_optimization&oldid=938811617}</source>
  <pubdate>2020</pubdate>
  <note>[Online; accessed 4-February-2020]</note>
</bibl>

</refgrp>
} 

\section*{Competing interests}
The authors declare that they have no competing interests.

\section*{Author's contributions}
The authors contributed equally to this work. All authors read and approved the final manuscript.

\section*{Acknowledgements}
We thank Leonardo Ermann for useful discussions. We thank UN Comtrade for providing to us a friendly access to their database.

\section*{Funding}
Programme Investissements d’Avenir ANR-15-IDEX-0003, ISITE-BFC (GNETWORKS project);
Bourgogne Franche-Comté region (APEX project);
ANR-11-IDEX-0002-02, reference ANR-10-LABX-0037-NEXT France (project THETRACOM).



%




\newpage

\setcounter{figure}{0}
\renewcommand{\thefigure}{A\arabic{figure}}
\setcounter{table}{0}
\renewcommand{\thetable}{A\arabic{table}}
\section*{Additional Files}

\subsection*{Additional file 1 --- Fraction of bankrupted countries for the WTN of 2004, 2008, 2012, and 2016 (model B)}
\label{add1}
\begin{figure}[h!]
	\centering
	\includegraphics[width=0.95\columnwidth]{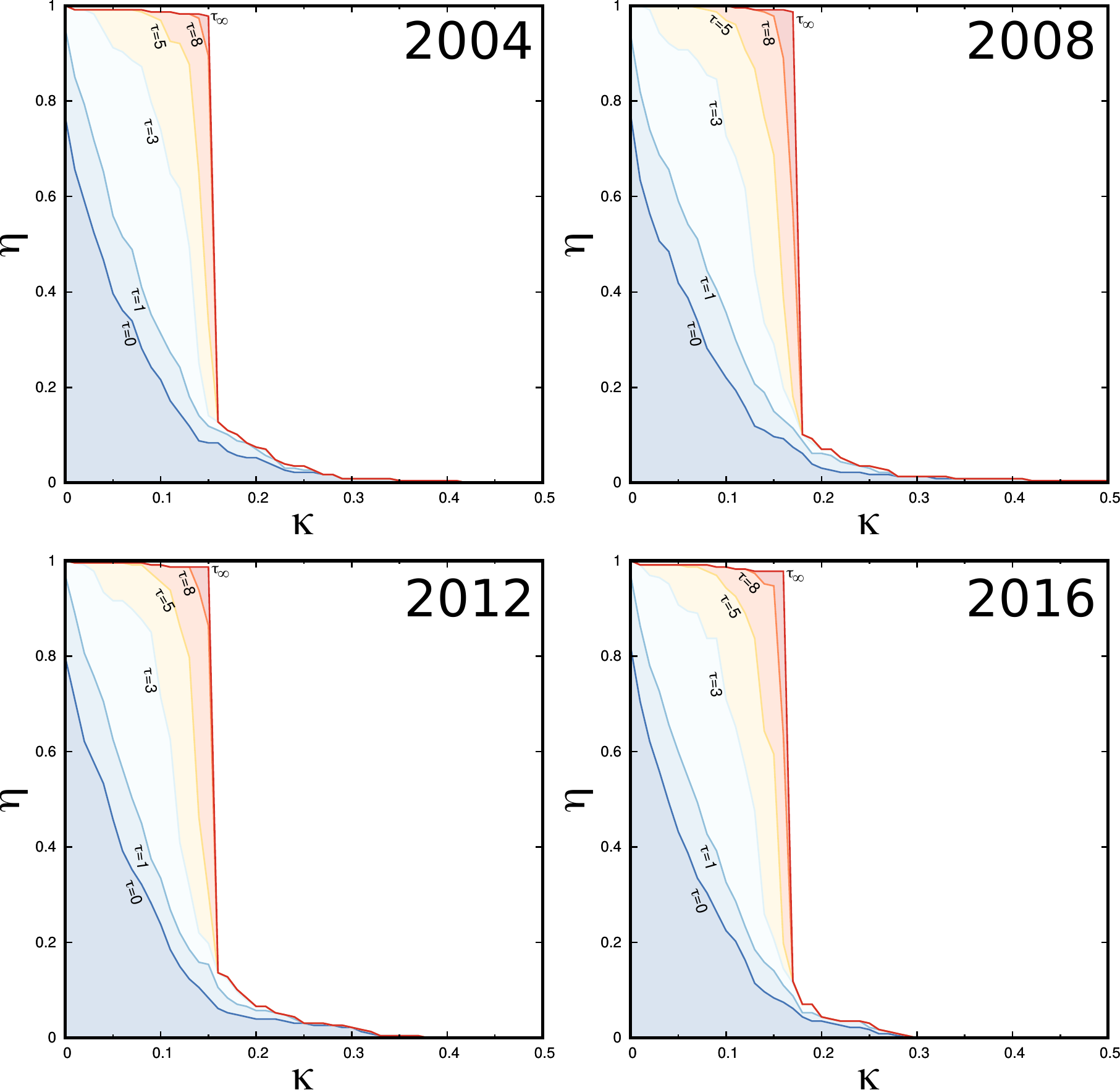}
	\caption{\csentence{Fraction of bankrupted countries for the WTN of 2004, 2008, 2012, and 2016.} Fraction $\eta$ of countries went to bankruptcy up to the $\tau$th stage of crisis contagion as a function of the bankruptcy threshold $\kappa$. The crisis contagion has been computed for the WTN of
		2004 (top left),
		2008 (top right),
		2012 (bottom left), and
		2016 (bottom right).
		Once a country goes to bankruptcy, it is prevented to import any product (model B). The damping factor is $\alpha=0.5$.}
	\label{Sfig1}
\end{figure}

\newpage
\subsection*{Additional file 2 --- List of the 38 countries remaining safe at $\tau_{\infty}$ for $\kappa = 0.1$ in 2004 (model A)}

\begin{table}[h!]
	\caption{List of the 38 countries remaining safe at $\tau_{\infty}$ for $\kappa = 0.1$ in 2004 (model A). The CC and SC columns give the ISO 3166-2 codes of the country and of its Sovereign country, respectively. The LT column gives the land type of the country, i.e. either Mainland (M) or Island (I). The three last columns give the percentages of gas and petroleum exported by the country. Countries are sorted primarily by the proportion of gas and petroleum in their exports (last column) and secondarily by the export total amount (fifth column).}
	\begin{tabular}{|>{\ttfamily}c|>{\ttfamily}c|c|p{10em}|r|r|r|r|}
		\hline
		\multicolumn{4}{|l|}{Country}&\multicolumn{4}{c|}{Exportation}\\
		\hline
		\multicolumn{1}{|c|}{CC}&\multicolumn{1}{c|}{SC}&\multicolumn{1}{c|}{LT}&Name&\multicolumn{1}{p{4em}|}{\centering Total \\($10^6$USD)}&\multicolumn{1}{p{4em}|}{\centering Gas\\(\permil)}&\multicolumn{1}{p{4em}|}{\centering Petroleum\\ (\permil)}&\multicolumn{1}{p{4em}|}{\centering Gas \& petroleum\\ (\permil)}\\
		\hline
		NG & NG & M & Nigeria & 34582.35 & 49.45 & 910.87 & 960.32\\
		SA & SA & M & Saudi Arabia & 110667.14 & 34.88 & 787.04 & 821.92\\
		RU & RU & M & Russia & 225850.70 & 51.36 & 447.20 & 498.56\\
		NR & NR & I & Nauru & 17.58 & 28.38 & 425.38 & 453.76\\
		PN & UK & I & Pitcairn Islands & 20.02 & 0.00 & 448.97 & 448.97\\
		TL & TL & I & East Timor & 155.31 & 21.89 & 274.94 & 296.83\\
		SC & SC & I & Seychelles & 520.47 & 0.02 & 194.06 & 194.08\\
		GU & US & I & Guam & 68.83 & 1.60 & 167.09 & 168.69\\
		KE & KE & M & Kenya & 3507.94 & 0.65 & 139.40 & 140.05\\
		BM & UK & I & Bermuda & 189.97 & 0.00 & 74.22 & 74.22\\
		MH & MH & I & Marshall Islands & 651.83 & 0.00 & 57.53 & 57.53\\
		KY & UK & I & Cayman Islands & 674.49 & 0.00 & 46.41 & 46.41\\
		VA & VA & M & Vatican & 3.19 & 0.00 & 18.58 & 18.58\\
		MP & US & I & Northern Mariana Islands & 18.97 & 0.00 & 10.81 & 10.81\\
		SH & UK & I & Saint Helena, Ascension and Tristan da Cunha & 15.52 & 0.00 & 9.99 & 9.99\\
		TC & UK & I & Turks and Caicos Islands & 30.78 & 9.52 & 0.01 & 9.53\\
		AS & US & I & American Samoa & 22.63 & 0.00 & 8.62 & 8.62\\
		FK & UK & I & Falkland Islands & 136.20 & 0.00 & 4.53 & 4.53\\
		IO & UK & I & British Indian Ocean Territory & 3.23 & 0.00 & 3.07 & 3.07\\
		TV & TV & I & Tuvalu & 2.12 & 1.95 & 0.01 & 1.95\\
		TK & NZ & I & Tokelau & 20.41 & 0.00 & 1.48 & 1.48\\
		SM & SM & M & San Marino & 53.24 & 0.00 & 0.92 & 0.92\\
		SB & SB & I & Solomon Islands & 193.42 & 0.00 & 0.84 & 0.84\\
		GL & DK & M & Greenland & 538.59 & 0.00 & 0.26 & 0.26\\
		PW & PW & I & Palau & 22.80 & 0.00 & 0.10 & 0.10\\
		CK & CK & I & Cook Islands & 17.01 & 0.00 & 0.04 & 0.04\\
		BT & BT & M & Bhutan & 57.88 & 0.00 & 0.01 & 0.01\\
		UM & US & I & United States Minor Outlying Islands & 33.55 & 0.00 & 0.00 & 0.00\\
		CX & AU & I & Christmas Island & 15.54 & 0.00 & 0.00 & 0.00\\
		PM & FR & I & Saint Pierre and Miquelon & 7.80 & 0.00 & 0.00 & 0.00\\
		CC & AU & I & Cocos (Keeling) Islands & 6.05 & 0.00 & 0.00 & 0.00\\
		NF & AU & I & Norfolk Island & 4.01 & 0.00 & 0.00 & 0.00\\
		GS & UK & I & South Georgia and the South Sandwich Islands & 3.50 & 0.00 & 0.00 & 0.00\\
		EH &  & M & Western Sahara & 2.11 & 0.00 & 0.00 & 0.00\\
		AQ &  & M & Antarctica & 2.05 & 0.00 & 0.00 & 0.00\\
		NU & NU & I & Niue & 1.88 & 0.00 & 0.00 & 0.00\\
		HM & AU & I & Heard and McDonald Islands & 0.95 & 0.00 & 0.00 & 0.00\\
		BV & NO & I & Bouvet Island & 0.25 & 0.00 & 0.00 & 0.00\\
		\hline
	\end{tabular}
	\label{table:2004-safe-0_1}
\end{table}

\newpage
\subsection*{Additional file 3 --- List of the 38 countries remaining safe at $\tau_{\infty}$ for $\kappa = 0.1$ in 2008 (model A)}

\begin{table}[h!]
	\caption{List of the 38 countries remaining safe at $\tau_{\infty}$ for $\kappa = 0.1$ in 2008 (model A). The CC and SC columns give the ISO 3166-2 codes of the country and of its Sovereign country, respectively. The LT column gives the land type of the country, i.e. either Mainland (M) or Island (I). The three last columns give the percentages of gas and petroleum exported by the country.
		Countries are sorted primarily by the proportion of gas and petroleum in their exports (last column) and secondarily by the export total amount (fifth column).}
	\begin{tabular}{|>{\ttfamily}c|>{\ttfamily}c|c|p{10em}|r|r|r|r|}
		\hline
		\multicolumn{4}{|l|}{Country}&\multicolumn{4}{c|}{Exportation}\\
		\hline
		\multicolumn{1}{|c|}{CC}&\multicolumn{1}{c|}{SC}&\multicolumn{1}{c|}{LT}&Name&\multicolumn{1}{p{4em}|}{\centering Total \\($10^6$USD)}&\multicolumn{1}{p{4em}|}{\centering Gas\\(\permil)}&\multicolumn{1}{p{4em}|}{\centering Petroleum\\ (\permil)}&\multicolumn{1}{p{4em}|}{\centering Gas \& petroleum\\ (\permil)}\\
		\hline
		TL & TL & I & East Timor & 169.40 & 816.28 & 0.28 & 816.56\\
		RU & RU & M & Russia & 570605.26 & 44.12 & 550.37 & 594.49\\
		BV & NO & I & Bouvet Island & 41.07 & 0.00 & 357.29 & 357.29\\
		CK & CK & I & Cook Islands & 33.11 & 0.00 & 239.31 & 239.31\\
		BM & UK & I & Bermuda & 1849.83 & 3.54 & 104.77 & 108.30\\
		UM & US & I & United States Minor Outlying Islands & 24.16 & 0.00 & 68.68 & 68.68\\
		SZ & SZL & M & Eswatini & 1058.15 & 0.00 & 61.57 & 61.57\\
		GS & UK & I & South Georgia and the South Sandwich Islands & 1.62 & 0.00 & 21.96 & 21.96\\
		AD & AD & M & Andorra & 175.92 & 7.70 & 9.86 & 17.56\\
		NF & AU & I & Norfolk Island & 4.04 & 0.00 & 15.57 & 15.57\\
		EH &  & M & Western Sahara & 11.31 & 0.00 & 14.72 & 14.72\\
		TV & TV & I & Tuvalu & 4.02 & 0.00 & 11.10 & 11.10\\
		SH & UK & I & Saint Helena, Ascension and Tristan da Cunha & 43.99 & 0.00 & 3.38 & 3.38\\
		NU & NU & I & Niue & 9.67 & 0.00 & 2.77 & 2.77\\
		GU & US & I & Guam & 77.22 & 0.00 & 2.33 & 2.33\\
		AQ &  & M & Antarctica & 2.55 & 0.00 & 1.75 & 1.75\\
		GL & DK & M & Greenland & 753.16 & 0.00 & 1.38 & 1.38\\
		WS & WS & I & Samoa & 89.95 & 0.00 & 1.28 & 1.28\\
		WF & FR & I & Wallis and Futuna & 18.64 & 0.00 & 0.98 & 0.98\\
		VU & VU & I & Vanuatu & 568.61 & 0.00 & 0.67 & 0.67\\
		LS & LS & M & Lesotho & 873.69 & 0.27 & 0.39 & 0.66\\
		KI & KI & I & Kiribati & 14.02 & 0.00 & 0.47 & 0.47\\
		NR & NR & I & Nauru & 126.51 & 0.00 & 0.33 & 0.33\\
		FM & FM & I & Federated States of Micronesia & 28.57 & 0.00 & 0.13 & 0.13\\
		AS & US & I & American Samoa & 70.09 & 0.00 & 0.08 & 0.08\\
		BT & BT & M & Bhutan & 688.82 & 0.00 & 0.03 & 0.03\\
		IO & UK & I & British Indian Ocean Territory & 8.25 & 0.00 & 0.02 & 0.02\\
		SB & SB & I & Solomon Islands & 383.50 & 0.00 & 0.01 & 0.01\\
		PW & PW & I & Palau & 29.04 & 0.00 & 0.01 & 0.01\\
		FK & UK & I & Falkland Islands & 196.72 & 0.00 & 0.00 & 0.00\\
		GW & GW & M & Guinea-Bissau & 135.22 & 0.00 & 0.00 & 0.00\\
		CX & AU & I & Christmas Island & 52.61 & 0.00 & 0.00 & 0.00\\
		CC & AU & I & Cocos (Keeling) Islands & 29.68 & 0.00 & 0.00 & 0.00\\
		PM & FR & I & Saint Pierre and Miquelon & 17.42 & 0.00 & 0.00 & 0.00\\
		MP & US & I & Northern Mariana Islands & 12.64 & 0.00 & 0.00 & 0.00\\
		PN & UK & I & Pitcairn Islands & 9.38 & 0.00 & 0.00 & 0.00\\
		VA & VA & M & Vatican & 2.52 & 0.00 & 0.00 & 0.00\\
		HM & AU & I & Heard and McDonald Islands & 0.53 & 0.00 & 0.00 & 0.00\\
		\hline
	\end{tabular}
	\label{table:2008-safe-0_1}
\end{table}

\newpage
\subsection*{Additional file 4 --- List of the 32 countries remaining safe at $\tau_{\infty}$ for $\kappa = 0.1$ in 2012 (model A)}

\begin{table}[h!]
	\caption{List of the 32 countries remaining safe at $\tau_{\infty}$ for $\kappa = 0.1$ in 2012 (model A). The CC and SC columns give the ISO 3166-2 codes of the country and of its Sovereign country, respectively. The LT column gives the land type of the country, i.e. either Mainland (M) or Island (I). The three last columns give the percentages of gas and petroleum exported by the country.
		Countries are sorted primarily by the proportion of gas and petroleum in their exports (last column) and secondarily by the export total amount (fifth column).}
	\begin{tabular}{|>{\ttfamily}c|>{\ttfamily}c|c|p{10em}|r|r|r|r|}
		\hline
		\multicolumn{4}{|l|}{Country}&\multicolumn{4}{c|}{Exportation}\\
		\hline
		\multicolumn{1}{|c|}{CC}&\multicolumn{1}{c|}{SC}&\multicolumn{1}{c|}{LT}&Name&\multicolumn{1}{p{4em}|}{\centering Total \\($10^6$USD)}&\multicolumn{1}{p{4em}|}{\centering Gas\\(\permil)}&\multicolumn{1}{p{4em}|}{\centering Petroleum\\ (\permil)}&\multicolumn{1}{p{4em}|}{\centering Gas \& petroleum\\ (\permil)}\\
		\hline
		RU & RU & M & Russia & 640181.69 & 82.16 & 565.46 & 647.63\\
		TC & UK & I & Turks and Caicos Islands & 94.84 & 0.00 & 585.21 & 585.21\\
		GU & US & I & Guam & 146.08 & 7.94 & 519.57 & 527.51\\
		AS & US & I & American Samoa & 91.34 & 6.58 & 379.91 & 386.49\\
		BV & NO & I & Bouvet Island & 55.78 & 0.00 & 288.63 & 288.63\\
		AQ &  & M & Antarctica & 150.45 & 0.00 & 243.27 & 243.27\\
		KY & UK & I & Cayman Islands & 622.35 & 0.00 & 19.92 & 19.92\\
		HM & AU & I & Heard and McDonald Islands & 245.23 & 0.00 & 11.25 & 11.25\\
		VA & VA & M & Vatican & 7.71 & 0.00 & 1.45 & 1.45\\
		NU & NU & I & Niue & 3.59 & 0.00 & 1.17 & 1.17\\
		SH & UK & I & Saint Helena, Ascension and Tristan da Cunha & 19.12 & 0.00 & 0.99 & 0.99\\
		MP & US & I & Northern Mariana Islands & 3.65 & 0.01 & 0.86 & 0.86\\
		NR & NR & I & Nauru & 100.46 & 0.00 & 0.58 & 0.58\\
		GS & UK & I & South Georgia and the South Sandwich Islands & 4.03 & 0.00 & 0.46 & 0.46\\
		SB & SB & I & Solomon Islands & 572.38 & 0.00 & 0.40 & 0.40\\
		YT & FR & I & Mayotte & 28.29 & 0.00 & 0.16 & 0.16\\
		AI & UK & I & Anguilla & 10.29 & 0.00 & 0.15 & 0.15\\
		CK & CK & I & Cook Islands & 51.41 & 0.00 & 0.14 & 0.14\\
		FO & DK & I & Faroe Islands & 1001.96 & 0.00 & 0.10 & 0.10\\
		CX & AU & I & Christmas Island & 38.71 & 0.00 & 0.05 & 0.05\\
		IO & UK & I & British Indian Ocean Territory & 32.05 & 0.00 & 0.05 & 0.05\\
		UM & US & I & United States Minor Outlying Islands & 20.42 & 0.00 & 0.03 & 0.03\\
		TK & NZ & I & Tokelau & 43.62 & 0.00 & 0.03 & 0.03\\
		VU & VU & I & Vanuatu & 454.29 & 0.00 & 0.01 & 0.01\\
		FK & UK & I & Falkland Islands & 210.46 & 0.01 & 0.00 & 0.01\\
		SM & SM & M & San Marino & 128.83 & 0.00 & 0.00 & 0.00\\
		ER & ER & M & Eritrea & 47.89 & 0.00 & 0.00 & 0.00\\
		PM & FR & I & Saint Pierre and Miquelon & 7.12 & 0.00 & 0.00 & 0.00\\
		CC & AU & I & Cocos (Keeling) Islands & 7.05 & 0.00 & 0.00 & 0.00\\
		PN & UK & I & Pitcairn Islands & 6.71 & 0.00 & 0.00 & 0.00\\
		NF & AU & I & Norfolk Island & 3.85 & 0.00 & 0.00 & 0.00\\
		WF & FR & I & Wallis and Futuna & 1.30 & 0.00 & 0.00 & 0.00\\
		\hline
	\end{tabular}
	\label{table:2012-safe-0_1}
\end{table}

\newpage
\subsection*{Additional file 5 --- List of the 11 countries remaining safe at $\tau_{\infty}$ for $\kappa = 0.1$ in 2016 (model A)}

\begin{table}[h!]
	\caption{List of the 11 countries remaining safe at $\tau_{\infty}$ for $\kappa = 0.1$ in 2016 (model A). The CC and SC columns give the ISO 3166-2 codes of the country and of its Sovereign country, respectively. The LT column gives the land type of the country, i.e. either Mainland (M) or Island (I). The three last columns give the percentages of gas and petroleum exported by the country.
		Countries are sorted primarily by the proportion of gas and petroleum in their exports (last column) and secondarily by the export total amount (fifth column).}
	\begin{tabular}{|>{\ttfamily}c|>{\ttfamily}c|c|p{10em}|r|r|r|r|}
		\hline
		\multicolumn{4}{|l|}{Country}&\multicolumn{4}{c|}{Exportation}\\
		\hline
		\multicolumn{1}{|c|}{CC}&\multicolumn{1}{c|}{SC}&\multicolumn{1}{c|}{LT}&Name&\multicolumn{1}{p{4em}|}{\centering Total \\($10^6$USD)}&\multicolumn{1}{p{4em}|}{\centering Gas\\(\permil)}&\multicolumn{1}{p{4em}|}{\centering Petroleum\\ (\permil)}&\multicolumn{1}{p{4em}|}{\centering Gas \& petroleum\\ (\permil)}\\
		\hline
		GS & UK & I & South Georgia and the South Sandwich Islands & 0.34 & 0.00 & 83.01 & 83.01\\
		AQ &  & M & Antarctica & 10.28 & 0.00 & 11.67 & 11.67\\
		NU & NU & I & Niue & 2.30 & 0.00 & 10.50 & 10.50\\
		FK & UK & I & Falkland Islands & 257.30 & 0.00 & 5.87 & 5.87\\
		CC & AU & I & Cocos (Keeling) Islands & 4.59 & 0.00 & 0.29 & 0.29\\
		EH &  & M & Western Sahara & 8.92 & 0.00 & 0.01 & 0.01\\
		BV & NO & I & Bouvet Island & 0.86 & 0.00 & 0.00 & 0.00\\
		IO & UK & I & British Indian Ocean Territory & 20.16 & 0.00 & 0.00 & 0.00\\
		SH & UK & I & Saint Helena, Ascension and Tristan da Cunha & 26.64 & 0.00 & 0.00 & 0.00\\
		PN & UK & I & Pitcairn Islands & 1.29 & 0.00 & 0.00 & 0.00\\
		HM & AU & I & Heard and McDonald Islands & 0.12 & 0.00 & 0.00 & 0.00\\
		\hline
	\end{tabular}
	\label{table:2016-safe-0_1}
\end{table}

\newpage
\subsection*{Additional file 6 --- Evolution of the crisis contagion in the 2016 WTN}
\label{sec:add6}
See also the video at \url{http://perso.utinam.cnrs.fr/~lages/datasets/WTNcrisis/}

\newpage
\subsection*{Additional file 7 --- Crisis contagion network for years 2004, 2008, 2012, and 2016}
\begin{figure*}[h!]
	\centering
	\includegraphics[width=0.7\columnwidth]{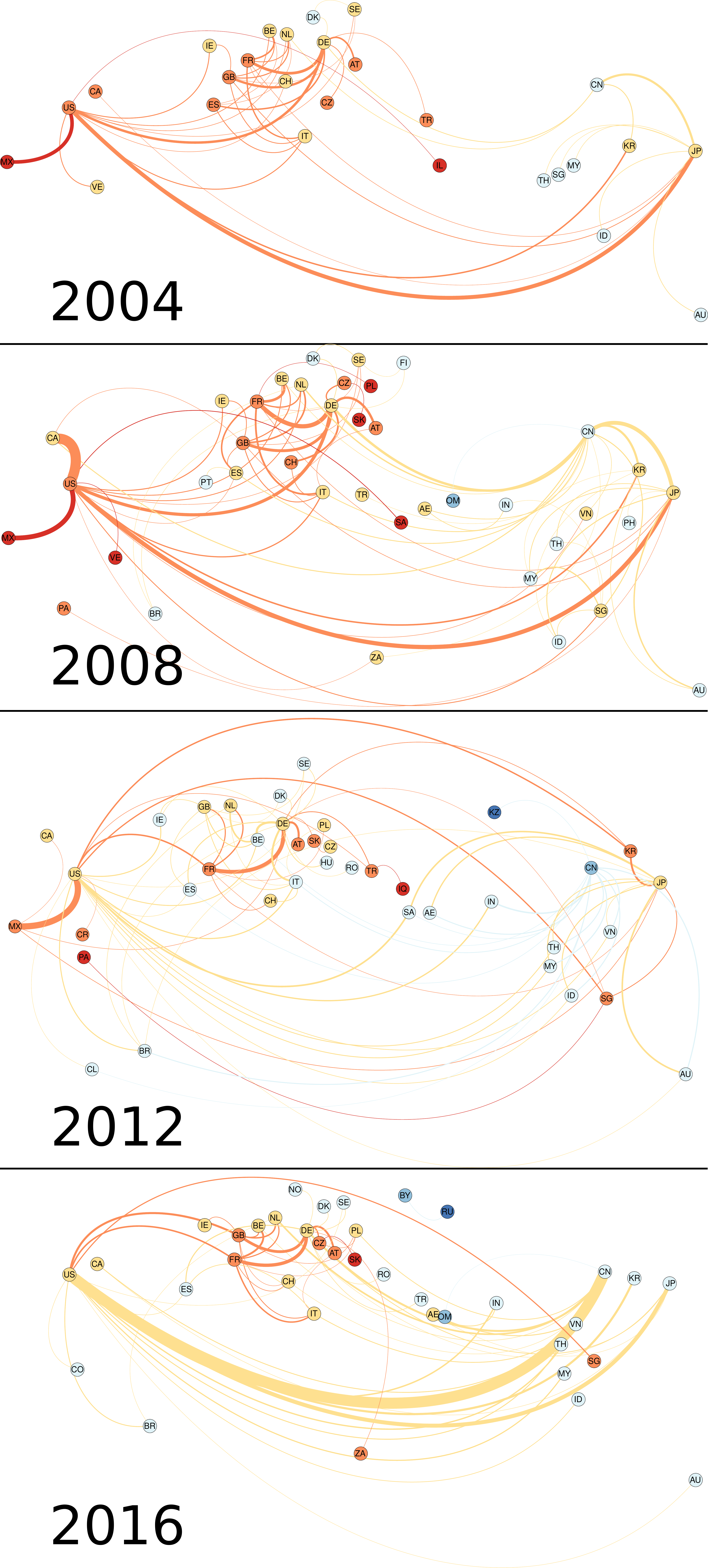}
	\caption{\csentence{Crisis contagion network for years 2004, 2008, 2012, and 2016.} Same as Figure \ref{fig8} but only exports greater than $10^{10}$ USD are represented.}
	\label{Sfig2}
\end{figure*}

%

\end{backmatter}
\end{document}